\documentclass[12pt]{olplainarticle}

\usepackage{amsmath}
\usepackage{amsthm}
\usepackage{amssymb}
\usepackage{nicefrac}
\usepackage{comment}
\usepackage{hyperref}

\usepackage{algorithm}
\usepackage[indLines=true]{algpseudocodex}

\theoremstyle{plain}
\newtheorem{thm}{Theorem}[section]
\newtheorem{lem}[thm]{Lemma}

\theoremstyle{definition}

\theoremstyle{remark}

\newcommand{\A}{\mathbb A}

\newcommand{\R}{\mathbb R}
\newcommand{\Rd}{\R^d}
\newcommand{\bbS}{\mathbb S}
\newcommand{\dd}{\mathrm d}

\newcommand{\cN}{\mathcal N}

\newcommand{\cV}{\mathcal V}
\newcommand{\cM}{\mathcal M}

\newcommand{\cB}{\mathcal B}

\newcommand{\cU}{\mathcal U}

\newcommand{\candidate}{r}
\newcommand{\node}{x}
\newcommand{\direction}{u}

\newcommand{\bsigma}{\boldsymbol \sigma}
\newcommand{\bfn}{{\bf n}}
\newcommand{\br}{\boldsymbol r}
\newcommand{\bfeta}{{\boldsymbol \eta}}
\newcommand{\ups}{\upsilon}

\usepackage{xcolor}

\begin{document}

\title{Voronoi Graph - Improved raycasting and integration schemes for high dimensional Voronoi diagrams}

\author[1]{Alexander Sikorski}
\author[2]{Martin Heida}
\affil[1]{Freie Universität Berlin, Zuse Institute Berlin}
\affil[2]{Weierstrass Institute for Applied Analysis and Stochastics}


\begin{abstract}
The computation of Voronoi Diagrams, or their dual Delauney triangulations is difficult in high dimensions. In a recent publication Polianskii and Pokorny propose an iterative randomized algorithm facilitating the approximation of Voronoi tesselations in high dimensions.

In this paper, we provide an improved vertex search method that is not only exact but even faster than the bisection method that was previously recommended.

Building on this we also provide a depth-first graph-traversal algorithm which allows us to compute the entire Voronoi diagram. 
This enables us to compare the outcomes with those of classical algorithms like qHull, which we either match or marginally beat in terms of computation time.

We furthermore show how the raycasting algorithm naturally lends to a Monte Carlo approximation for the volume and boundary integrals of the Voronoi cells, both of which are of importance for finite Volume methods. We compare the Monte-Carlo methods to the exact polygonal integration, as well as a hybrid approximation scheme.
\end{abstract}

\flushbottom
\maketitle
\thispagestyle{empty}
\newcommand{\raycast}{\textsc{RayCast }}
\section{Introduction}

Since they have been discovered in the first half of the 20th century, Voronoi Diagrams \cite{voronoi1908nouvelles} and Delaunay triangulations \cite{delaunay1934sphere} have become fundamental cornerstones in computational geometry and computational sciences.
They are often used for clustering or mesh generation and find applications in many fields such as physics, biology, astronomy as well as archeology, physiology and economics \cite{okabe2009spatial}.

Several methods were suggested for their calculation such as the Bowyer-Watson algorithm \cite{bowyer1981computing,watson1981computing} to directly compute the Voronoi Diagram or the projection in higher dimension and using the quickhull algorithm  \cite{quickhull} in order to calculate the Delaunay triangulation.



The presented work is one out of a series of two papers. It builds heavily upon the Voronoi Graph algorithm by Polianskii and Pokorny \cite{polianskii2019voronoi,polianskii2020voronoi}.
The core of the algorithm relies on a raycasting procedure computing the first intersection of a given ray with the Voronoi cells' boundaries based on a nearest neighbour search (NN).
In the original version the authors use a continuous binary search to approximate the intersection point whose precision depends on the number of NN evaluations.
We replace this subroutine by a deterministic version that terminates with only a few NN evaluations which is not only faster but also exact.
We furthermore show how the random search can be modified to obtain an exhaustive search returning the exact Voronoi diagram whilst avoiding the recomputation of already discovered vertices and compare its runtime with the classical qHull algorithm.
Note however, that unlike qHull this is an iterative algorithm, which in turn can be used to compute approximate or local Voronoi diagrams as well.

The algorithm performs with $N\ln(N)$ for $N$ iid distributed nodes and hence scales as good as the Bowyer-Watson algorithm. However, to our knowledge all previous works require the input nodes to be in general position, i.e. there is no point in $\Rd$ with more than $d+1$ nearest neighbors at once. 
In a companion paper \cite{heida2023voronoi} it is shown how this algorithm can be generalized to any set of points therefore also facilitating the computation of degenerate Voronoi diagrams.
It furthermore generalizes the algorithm from this article to be applied to polytopal domains or periodic diagrams. Wheras that article is very mathematical, particularly due to the treatmend of nodes in non-general positions, the current work focuses on the fundamental idea of the raycasting algorithm.

Based on raycasting approach we furthermore introduce several methods to calculate or approximate the volume of Voronoi cells and the area of interfaces between cells as well as several methods to approximate integrals of functions over cells or interfaces.



The improved \raycast, exhaustive search and Monte-Carlo integration where initially implemented in the lightweight Julia package \verb|VoronoiGraph.jl|\cite{VoronoiGraph} and in this article we will follow this more simplistic approach for didactic reasons.
These ideas were picked up and extended in \verb|HighVoronoi.jl|\cite{HighVoronoi} providing exhaustive documentation, support for points in non-generic position, boundary conditions, the polygonal and heuristic integration rules and more, which are described in more detail in the accompanying publication \cite{heida2023voronoi}.

\subsection*{Outline}

In section \ref{sec:Voronoi-Algo} we outline the idea of calculating the Voronoi Diagram using the raycasting approach, we introduce our new incircle \raycast procedure and provide the exhaustive graph traversal to compute the entire Voronoi diagram.

In Section \ref{sec:Integration} we will introduce two different methods to calculate cell volumes as well as interface areas and three different ways to integrate functions over cells and interfaces numerically. We will introduce the algorithms and also briefly discuss their respective mathematical background.

Finally, in Section \ref{sec:performance} we will study the performance as well as the accuracy of our proposed methods. We compare the performance of the incircle \raycast to its bisection predecessor and the compute time for the entire Voronoi diagram of our implementations to qHull. We conclude with a numerical comparison of the integration routines.

\section{Voronoi diagrams via raycasting}\label{sec:Voronoi-Algo}

\begin{figure}
    \centering
    \includegraphics[width=0.5 \textwidth]{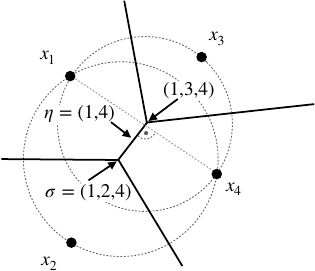}
    \caption{Illustration of a voronoi diagram with labeled generators, two vertices and their connecting edge. The circles illustrate the equidistance of the generators to the respective vertices. The edge is equidistant to  its generators as well as perpendicular to the plane spanned by them.
    }
     
    \label{fig:voronoi}
\end{figure}

\subsection{Notation}

We start by collecting a few key terms and notations. For a more comprehensive introduction and summary of the geometric aspects in the context of this algorithm, we recommend the work by \cite{polianskii2020voronoi}.

As mentioned in the introduction, we assume that the \emph{nodes} $X=(x_1,\dots,x_N)$ are in general position.
This implies that a \emph{vertex} $\nu$ is defined by $d+1$ nodes, which we also call \emph{generators}, all being the nearest neighbor to $\nu$ at a common distance.
We refer to the set of all Voronoi vertices as $\cV$ and the Voronoi \emph{cell} generated by $x_i \in X$ as $C_i$.
The Voronoi diagram is dual to the Delauney complex in the sense that every $k-face$ of a Voronoi simplex is dual to a $d-k$ simplex of the Delauney complex. , e.g. a ($d$-dimensional) cell $C_i$ is dual to the $0$-simplex $X_i$, a ($0$-dimensional) vertex is dual to the $d$-simplex spanned by its $d+1$ generators and an ($1$-dimensinal) edge is dual to the $d-1$ simplex spanned by the $d$ generators surrounding the edge.
In general we will denote by $\sigma \subset \mathcal P(\{1,...,N\})$ the indices to a set of generators and by $X_\sigma := \{ X_i | i\in \sigma\}$ the set of the corresponding generators.
Due to the duality we will also sometimes refer to a vertex, an edge, etc. directly in terms of their generators $\sigma$.

In the context of the algorithm we store each vertex $\nu$ by the tuple of its generators and its coordinate, $(\sigma, r)$.


An edge emerging at a vertex $(\bsigma,r)$ can hence be identified with a subset $\bfeta\subset\bsigma$ of $d$ generators. The edge is characterized by starting at $r$ and pointing away from the single generator $x_i$, $\{i\}=\bsigma \setminus \bfeta$ and along the vector $u$  orthogonal to the hyper plane spanned by the generators $X_\bfeta$.
We call $E(\bsigma)\subset \bsigma$ the set of all $d+1$ edges $\bfeta$ emerging at the vertex $\bsigma$ and $E(\bsigma,i)\subset E(\bsigma)$ the set of all $d$ edges emerging at $\sigma$ and sharing the common generator $i\in\bfeta$.

Whilst usually each egde $\bfeta$ has two vertices $\bsigma_\bfeta^1$ and $\bsigma_\bfeta^2$, i.e. $\bfeta=\bsigma_\bfeta^1\cap\bsigma_\bfeta^2$ some edges of the Voronoi diagram will become unbounded and are thus specified by a single vertex and a direction only. These edges belong to cells that are unbounded themselves and  we denote the set of unbounded edges by $\cB^\infty$.

For convenience in the algorithm, we store all nodes $X$, vertices $\cV := \{\nu\} $, edges $E$ and unbounded edges $\cB^\infty$ in a single data structure, the \emph{mesh} $\cM$:
$$\cM=(X,\cV,E,\cB^\infty)\,.$$

\newcommand{\gencand}{x}

\subsection{The incircle \raycast procedure}
\label{sec:raycast}

\begin{figure}
    \centering
    \includegraphics[width=1\textwidth]{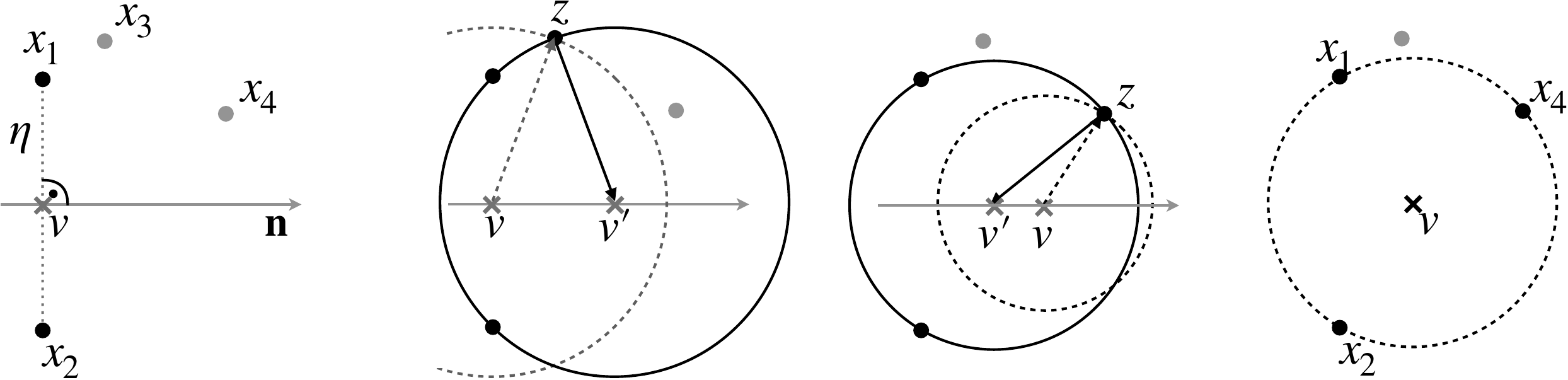}
    \caption{Illustration of the incircle \raycast search in 2D. From left to right:
    (a) We start with an initial guess $\candidate$ (equidistant to the current generators) and a search direction (perpendicular to the generator's plane) $\direction$
    (b) Nearest neighbour search (on the right half-plane) leads to a possible generator $\gencand$. The projection then leads to the vertex candidate $\candidate'$
    (c) Nearest neighbour search finds another possible generator, leading to another candidate.
    (d) Nearest neighbour search returns no other generator, confirming the incircle criterion.
    }
     
    \label{fig:raycast-figure}
\end{figure}

The fundamental algorithm for the computation of the Voronoi vertices is the \raycast procedure.
Here, we present a novel \raycast algorithm that improves upon its predecessor by requiring fewer nearest-neighbor evaluations and returning the exact vertex rather than an approximation.
It is based on the so-called \emph{incircle-criterion}, which states that a point $candidate$ is a vertex of the Voronoi diagram corresponding to $X$ if and only if a sphere surrounding $candidate$ contains at least $d+1$ generators but no additional generators within its interior.

The \raycast algorithm starts from candidate position $\candidate$ (usually a vertex) which is equidistant to a set of generators $X_\eta := \{X_i|i \in \eta\} \subset X$ specified by $\eta$ and a normalized search direction $\direction \in \Rd, ||\direction||=1$ that is orthogonal to the lower dimensional affine space spanned by $X_\eta$. 

In the following we describe the situation of searching for a vertex, i.e. where $|\eta| = d$ describes an edge and the candidate position consists of a known vertex' coordinates. 
However, this algorithm works for any $|\eta| \le d$ to find the respective next lower dimensional facet (e.g. a face given the whole cell in the case $|\eta|=1$).

Note that any vertex $r'$ having $\eta$ as subset of its generators must lay on the ray specified by $r+tu$ for some $t$.
Furthermore it must satisfy the incircle condition, i.e. there needs to be another generator $X_j\in X, j \not \in \eta$ such that $r'$ is equidistant to $X_\sigma$ and $X_j$ and there is no other generator.
We restrict our search to $t>0$, i.e. only into into the direction of $u$, for the uniqueness of the solution.

This is achieved in an iterative manner by looking for possible generators $\gencand$ of the desired vertex via a nearest neighbour search (restricted to the proper halfspace by $t>0$) around the candidate $\candidate$. Once a possible generator candidate $\gencand$ is found, the resulting vertex candidate 
\begin{equation}
\label{eq:candidate}
    \candidate' = \candidate + t \direction
\end{equation}  
has to be equidistant to $\gencand$ and a known generator $\node_0\in X_\eta$, which allows to compute its hypothetical position along the ray\footnote{Whilst this projection was already used in \cite{polianskii2020voronoi}, 
they did not take advantage of it in the way we do by starting the nearest neighbour search at the computed position}:
\begin{equation*}
    |\candidate'-\node_0|^2=|\candidate'-\gencand|^2
    \quad\Longleftrightarrow\quad
    |\candidate-\node_0|^2 + 2t \left< \direction,\candidate-\node_0\right>
    =|\candidate-\gencand|^2 + 2t \left< \direction, \candidate-\gencand \right>
\end{equation*}
which ultimately yields 
\begin{equation}
\label{eq:t}
    t = \frac{|\candidate - \node|^2 - |\candidate - \gencand_0|^2}{2 \left< \direction , (\gencand-\node_0)\right>}.
\end{equation}

We then continue iterating this procedure of nearest neighbour search and projection from the respectively current vertex candidate $r'$.
This continues until at some point no new generator is found which is closer to $\candidate'$ then the previously known ones. This means that the candidate indeed satisfies the incircle criterion and thus indeed is a vertex of the Voronoi diagram.

That this routine indeed terminates follows from the fact that the search radius is decreasing over the iterations and only a finite number of candidates is available.
For a mathematically rigorous formulation and proof of this argument, even for nodes in degenerate positions, we refer to \cite{heida2023voronoi}.

\begin{figure}
    \centering
    \includegraphics[width=0.3\textwidth]{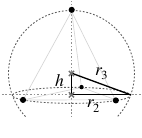}
    \caption{Schematic of the construction of the center of a regular 3--simplex from the regular 2--simplex center using Pythagoras' theorem.
    }
    \label{fig:heuristic}
\end{figure}

The search performance can be improved by a good initial guess. If the start point for the search was the desired target vertex the algorithm would terminate immediately.
For a heuristic, let us assume that the generators of the desired target vertex form a regular simplex, i.e. they are equidistant to each other.
We can then use the relation between the circumsphere's radius $r_d$ and the edge length $l$ in for a regular $d-$simplex,
\begin{equation}
    r_d = l\sqrt{\frac{d}{2(d+1)}},
\end{equation}
to construct the vertex position.
Using the Pythagoras' theorem (see Figure \ref{fig:heuristic}), $h_d^2 + r_{d-1}^2 = r_d^2$, and solving for the height $h_d$ of the new vertex, i.e. the distance between the center of the $d-1$-simplex and the $d$-simplex we arrive at 
\begin{equation}
    h_d=\frac{r_{d-1}}{\sqrt{(d+1)(d-1)}}.
\end{equation}

In Algorithm \ref{alg:raycast}, $\Call{InitialHeuristic}{}$ we shift the candidate point onto the plane spanned by the $d$ prescribed generators $\sigma$ by an orthogonal projection, compute the radius $r_{d-1}$ as the distance of this point to a generator and walk length $h_d$ in the provided direction $u$.

\subsection{The VoronoiGraph algorithm}

Using the \raycast algorithm we are in the position to travel along the edges of the Voronoi Diagram and compute new vertices from old ones, as well as discovering them ab initio by shooting rays onto successively lower-dimensional faces, just as in the original work \cite{polianskii2020voronoi}.

In that work they also introduced a random walk algorithm that approximates the Voronoi Diagram. They rightfully argue that the complexity of Voronoi diagrams explodes in higher dimension (see Section \ref{sec:complexity}) and hence only approximate Voronoi diagrams can be used in that setting. Of course our incircle \raycast can be used in that setting as well, inheriting any increase in performance.
However, as we will see in Section \ref{sec:perfbench}, we can use the improved \raycast as well to compute the exact Voronoi diagrams with competitive performance.

To this end we modify the original random walk to perform an exhaustive search which successively visits all vertices' neighbours and thus explores the whole diagram.
We furthermore keep track of the edges in the sense that we count how many vertices of each edge we have visited so far. This does not only save walking any edge twice. As it may happen that two vertices belonging to a single edge can be explored from different position this makes the \raycast along that edge completely obsolete.

Ignoring boundary effects, in $d$ dimensions every vertex has $d+1$ edges and any edge belongs to 2 vertices, so that the number of edges equals $\frac{d+1}{2}$ times that of the vertices.
By keeping track of the edges we discover every vertex only once, resulting in $n$ calls to \raycast, as opposed to $n\frac{(d+1)}{2}$ when not keeping track of the edges.

The complete procedure is given in Algorithm \ref{alg:explore}.

Note that the dictionary keeping track of the edges may consume a lot of memory. In \cite{heida2023voronoi} the cells $C_i$ are explored in sequential order. This allows to reconstruct the edge counts for each cell individually and thereby reduces the memory footprint of storing the edges.

\newcommand{\EC}{G}
\newlength{\commentlen}
\setlength{\commentlen}{38ex}

\subsection{Remark on the computational complexity}
\label{sec:complexity}
As noted before the computation of the full Voronoi diagrams in high dimensions is an inherently hard problem.

In \cite{dwyer1993expected} a lower bound for the expectation value of the number of vertices per cell is provided in the limit for $n\rightarrow\infty$ uniformly distributed nodes $X$. This lower bound is proportional to the following constant for $k=d$ and suggesting a superexponential growth:

\begin{equation}\label{eqn:lowerfacetbound}
    C_{k,d} \ge \frac{2\pi ^ {k/2} d^{k-1}}{k(k+1)}
    \beta\left(\frac{dk}{2}, \frac{d-k+1}{2}\right)^{-1}
    \left(\frac{\Gamma(d/2)}{\Gamma((d+1)/2)}\right)^k
    .
\end{equation}
Here $\beta$ and $\Gamma$ denote the beta resp. gamma functions.
In the following table we computed the expected number of verticed for $d=1, ... 10$:
\begin{table}[h]
\centering
    \begin{tabular}{c|c|c|c|c|c|c|c|c|c|c}
    \textbf{d} & 2 & 3 & 4 & 5 & 6 & 7 & 8 & 9 & 10 \\
    \textbf{$\mathbb{E}[\#V_i]$}  & 6.76 & 31.8 & 187 & 1296 & 1.03$\times 10^4$ & 9.04$\times 10^4$ & 8.72$\times 10^5$ & 9.09$\times 10^6$ & 1.02$\times 10^8$
\end{tabular}
    \caption{Expected number of vertices per cell for uniformly random data}
    \label{tab:dwyer}
\end{table}

Using HighVoronoi.jl with periodic boundary conditions we arrive at close empirical numbers:

\begin{table}[h]
\centering
\begin{tabular}{r|c|c|c|c|c|c}
Dimension &  2 & 3 & 4 & 5 & 6 & 7 \\\hline
Vertices / Cell & 6.6 & 29.2 & 191 & 1250 & 9720 & 91900 \\
Neighbors / Cell & 6.6 & 16.6 & 43 & 95 & 210 & 480 \\
\end{tabular}   
\caption{Vertices and neighbors per cell computed using HighVoronoi.jl
\label{table:empscaling}
\cite{heida2023voronoi}}
\end{table}

As one can verify qualitatively, the increase of vertices in the above data goes with a factor $d+2$ from $d\rightsquigarrow d+1$, while the number of neighbors increases by a factor between 2.2 and 2.6.

In this regard the sheer amount of vertices to compute for an exact exact Voronoi diagram in high dimensions, somewhere around $d>10$, will prohibit a solution.

In special cases, i.e. when staying away from the limit of $n\rightarrow\infty$ uniform nodes, we are still able to compute the exact solution.
For example we can compute the diagram for $n=100$ nodes in dimension $d=9$ where the number of neighbours and hence also vertices is limited by the data. Note however, that in that case probably almost all cells are neighbouring each other at the boundary of the diagram. Whether such an computation makes any sense depends on the application.

However, our improved \raycast can still be used for the approximate voronoi graph computation suggested in \cite{polianskii2020voronoi} as well as for the Monte-Carlo estimators below.
As we will see below the full exploration routine also provides a conceptually simple but equiperformant alternative to state of the art methods as \emph{qHull} and it proved usefull for the development and testing of the (experimental) correctness of the proposed \raycast.

\section{Integration}\label{sec:Integration}

In this section we introduce three different approaches for handling area and volume integrals.

We start with the most general, a Monte-Carlo approximation using random rays, which allows for the estimation of the area and volume respectively functions $f$ over them.

We then show how to exactly compute the areas and volumes by recursively constructing adequate pyramids and computing their size with the Leibnitz rule, providing an improved algorithm for the calculation of the corresponding determinants. This in turn will lead to a method providing the exact integral of a piecewise linear interpolant of a given function.

Finally we combine both approaches to obtain the \emph{heuristic Monte-Carlo} method, that uses Monte-Carlo to estimate the area and computes the linear interpolants integrals based on that estimate.




\subsection{Monte-Carlo Integration}\label{sec:monte-int}
The basic idea of the Monte-Carlo method is to sample directions uniformly from the sphere and determine the intersection of these rays with the cell's bondary. By weighting the function evaluations at these positions via the change of variables we obtain a classical Monte-Carlo estimator for the integrals.

We start by sampling a random direction uniformly on the sphere $\cU(\bbS^{d-1})$. This can be achieved by 
renormalizing the draw from a multivariate Gaussian $\hat y \sim \mathcal{N}(0, I(d))$:
\begin{equation}
    y  \sim \cU(\bbS^{d-1})
\end{equation}
Using the \verb|\raycast| algorithm at $x_i$ in direction $y$ we obtain a $l_i(y)>0$ such that $x_i+y l_i(y)\in\partial C_i\cup\partial C_j$ for some $j$ and we have the continuous bijection
\begin{equation}\label{eq:l_i}
    \phi_i: S^{n-1} \rightarrow \partial C_i,\quad \phi_i(y) = x_i + y l_i(y)\,.
\end{equation}
We can compute the size of the infinitesimal area element hit by the ray in direction $y$ based on its distance to $x_i$ and its angle to $\bfn_{ij}=\frac{x_j-x_i}{|x_j-x_i|}$. According to the changes of variables formula the ratio between the boundary and sphere area in direction $y$ is given by
\begin{equation}\label{eq:dS-dA}
    \frac{\dd A}{\dd S} (y) = l_i^{d-1}(y)\,|\bfn_{ij}\cdot y|^{-1}
\end{equation}
Where $\dd A$ resp. $\dd S$ represent the surface measure of the cell boundary resp. the sphere.

This ratio allows us to estimate the surface integral of a function $f$ by Monte Carlo estimation with $N$ samples $(y_j)_{j=1,\dots,N} \sim \mathcal{U}(\bbS^{d-1})$ via
\begin{equation}\label{eq:mc-help1}
    \int_{\partial C_{i}} f(y)\, \dd A(y) = \int_{\bbS^{d-1}} f(\phi_i(y)) \frac{\dd A}{\dd S}(y) \,\dd S(y) \approx \sum_{j=1}^{N} \frac{S_{d-1}}{N} \frac{\dd A}{\dd S}(y_j) f(\phi_i(y_j))
\end{equation}
where $S_{d-1} = \frac{2\pi ^{\frac{d}{2}}}{\Gamma (\frac{d}{2})}$ 
is the surface area of the sphere $\bbS^{d-1}$.

Note that from the \raycast procedure we also obtain the neighbouring cell $C_j$ such that this sum naturally splits into the contributions of the different faces $C_i\cup C_j$ actually estimating the individual integrals $\int_{\partial C_i} f dA$.

This method naturally generalizes to volume integration by additionally sampling along each ray. Similar to before, let $(y_j,t_j) \sim \mathcal{U}(\bbS^{d-1}\times(0,1))$. Then 
\begin{equation}
    \psi_i: \bbS^{d-1} \times (0,1) \rightarrow V_i \\ \psi_i(y, t) = x_i + t\, y\, l_i(y)
\end{equation} 
is a bijection between our sample space and $C_i\setminus\{x_i\}$.

With the change of variables formula and $Vol_d=\nicefrac{S_{d-1}}{d}$ for the spheres volume we derive the infinitesimal volume
\begin{equation}
    m_i(y, t) = \frac{S_{d-1}}{d} t^{d-1} l_i(y)^d 
\end{equation}

so that 
\begin{equation}
     \int_{V_{i}} dV(x) = \int_{\bbS^{d-1} \times [0,1]} m_i(y, t) \dd y \dd t 
     \approx \sum_{j=1}^{N} \frac{S_{d-1}}{dN} l_i(y_j)^d
\end{equation}
\begin{equation}\label{eq:full_MC_integration}
     \int_{V_{i}} f(x) dV(x) = \int_{\bbS^{d-1} \times [0,1]} f(\psi_i(y, t)) m_i(y, t) \dd y \dd t 
     \approx \sum_{j=1}^{N} \frac{S_{d-1}}{dN} f(\psi_i(y_j, t_j)) t_j^{d-1} l_i(y_j)^d
\end{equation}

We provide the a routine computing area, volume, surface and volume integrals alltogether in Algorithm \ref{alg:montecarlo}.

Note that depending on the cost of evaluation $f$ relative to the cost of the \raycast calls the evaluation of the Monte-Carlo integrals can be adjusted. If \raycast is costly we can make use of a single ray for multiple volume integral evaluations (as done in the suggested algorithm). If on the other hand $f$ is costly we can integrate $f$ with few samples but reweight the resulting integral with more samples estimating only the area/volume. For example for the area we have $\int_{\partial C_i} dA \approx F_\delta \frac{A^*}{A}$ where $A^*$ is a more precise estimate of the area obtained from more samples without $f$ evaluations.

Note further that this method does not need prior knowledge of the Voronoi diagram since it relies on the \raycast method only and hence is applicable even to very high dimensions.
In the case where the whole Voronoi diagram is already pre-computed one can speed up the \raycast calls by running its nearest neighbour search over the known neighbours only.

\subsection{Leibnitz Integration}\label{sec:leibnitz-int}
The basic idea is to recursively decompose the volume of a convex $d$-dimensional polytope into pyramids over a lower-dimensional base surfaces. To be more concrete, consider a covex polytope $P$ with $x$ in the interior of $P$. If $X_P$ denotes the vertices of the polytope, we could take e.g. the centroid $x=|X_P|^{-1}\sum_{y\in X_P}y$. However, if $P$ is the Voronoi cell of $x_i\in X$ we simply chose $x=x_i$. 

We then observe that the boundary of $P$ can be decomposed into $K$ different $d-1$ dimensional polytopes $(\tilde P_i^1)_{i=1,\dots K}$ and the volume of $P$ decomposes into $K$ pyramids with base $\tilde P^1_i$ and apex $x$ and volume $\frac1d|\tilde P_i^1|\,\mathrm{dist}(x,\tilde P_i^1)$. 

On the other hand, we can calculate each $d-1$ dimensional mass $|\tilde P_i^1|$ in terms of its centroid $x_{(i)}^1$ and its $d-2$ dimensional ''boundaries'' $\tilde P_j^2$. This can be iterated up to the $d-(d-1)$ dimensional $\tilde P^{d-1}_k$ that is simply given by the distance of two vertices that span one edge.

That said, we can associate with the vertices $r_1$ and $r_2$ of the final edge as well as $x_{(i)}^1,\dots,x_{(j)}^{d-2}$ a $d$-dimensional tetrahedron and the volume of $P$ is simply the sum of all the volumes of these tetrahedrons by the above iterative argument.


We calculate the volumes of these tetrahedrons using the relation between volume and a determinant. However, the cost of calculating determinants grows with $d!\cdot d$.
The following approach of iterating the Leibnitz rule will lower these costs to a bit more than $2d^2$ (we do not provide explicit value here), i.e. much smaller than $d!\cdot d$ for $d>2$.
\subsubsection{The Leibnitz rule}

Suppose $\ups_1,\dots \ups_d\in\Rd$ and consider $\A=(\ups_1,\dots,\ups_d)$ be the matrix with columns $\ups_i$. If linearly independent, the vectors $\ups_k$ define a paralellotope with volume $|\det\A|$. Moreover the $d$-dimensional pyramid with apex $0$ and the base defined by the vectors $\ups_d$ has volume $\frac{1}{d!}|\det\A|$. Given $h$, the distance of $0$ to the plane defined by $\ups_1,\dots \ups_d$, we find that the $d-1$-dimensional area defined by $\ups_1,\dots \ups_d$ is $\frac{1}{h(d-1)!}|\det\A|$. 
In particular, calculating the  exact volume of such a pyramid or the area of the base  boils down to a calculation of of $\det\A$. 

We denote $\A_{1,j}$ the submatrix of $\A$ where the first $column$ and the $j$-th row have been deleted. Then we find according to the Leibnitz rule:
$$\det\A=\sum_j (-1)^{1+j}\det\A_{1,j}\,.$$
Now, if we write 
$$T^d_k=\{\{j_1,\dots,j_k\}\subset\{1,\dots,d\}\}$$ for the set of ordered subsets of $\{1,\dots,d\}$ with precisely $k$ elements, we define for $k\in\{1,\dots,d\}$ and $\tau\in T^d_k$ the matrix $\A_{k,\tau}$ which emerges from erasing the first $k$ columns and the rows $\tau$. Then  
\begin{equation}\label{eq:def-minors}
\det\A_{k,\tau}=\sum_{j=1}^d (-1)^{1+j}\det\A_{k+1,\tau\cup\{j\}}\,,\qquad\det\A_{d, \tau}=(\ups_d)_{j\not\in\tau}\,.
\end{equation}

Based on the above observations we interpret $\A$ as a data set storing the so called minors $\det\A_{k,\tau}$. It is important to observe that $\det\A_{k,\tau}$ depends only on $\ups_{k+1},\dots,\ups_d$. In particular we may formulate the following algorithm:

Suppose \verb|UpdateMinors|($\A,\ups_j,j$) has been called from $j=d$ downto $j=2$ and for $\ups_2,\dots \ups_{d}$ we finally obtain 
$$\det\A=\verb|UpdateMinors|(\A,\ups_1,1)$$

\subsubsection{Integration of a piecewise linearly interpolated function}

We now show how the volume integral of a linear function over a cone can be decomposed into a lower dimensional integral over a base surface and a pointwise evaluation at its apex.

Let $\tilde B\subset\R^{d-1}$ -- identifying it with $B=\tilde B\times\{0\}\subset\Rd$ -- and let $x_0=(0,\dots,0,1)\in\Rd$. Let $\tilde f: \tilde B\to\R$ be continuous, let $f_0\in\R$ and write $x=(\tilde x, x_d)$ for $x\in\Rd$, $\tilde x\in\R^{d-1}$, $x_d\in\R$. Let $C=\mathrm{conv}(B\cup\{x_0\})\subset\Rd$ be the cone with base $B$ and apex $x_0$ and define $f:C\to\R$ by $f(\tilde x,x_d)= (1-x_d)\tilde f(\tilde x)+x_d f_0$, the linear interpolation between the values of $\tilde f$ on $B$ and the value $f_0$ in $x_0$. Then we find
\begin{align}
\int_C f\,\dd x & = \int_0^1\int_{(1-x_d)\tilde B} (1-x_d)\tilde f(\tilde x) +x_df_0\,\dd\tilde x\,\dd x_d \nonumber \\
& = \int_0^1\int_{\tilde B} (1-x_d)^d\tilde f(\tilde x)\,\dd\tilde x\,\dd x_d +\int_0^1|\tilde B|(1-x_d)^{d-1} x_df_0\,\dd\tilde x\,\dd x_d\nonumber \\
& = \frac1d\left(\frac{d}{d+1}\int_{\tilde B} \tilde f(\tilde x)\,\dd\tilde x+\frac{1}{d+1}|\tilde B|f_0\right)\label{eq:leibniz-int}
\end{align}
This formula is by its nature iterative over dimensions of subsets as is implemented accordingly in steps 1.(f) and 2.(f) of the \verb|IterativeVolume| algorithm below. 

In particular, we can apply \eqref{eq:leibniz-int} when we perform the iteration outlined at the beginning of Section \ref{sec:leibnitz-int} evaluating $f_0$ in the centroids. By the iterative nature of \verb|IterativeVolume| this leads to the precise integral of a function $F$ that coincides with the original $f$ in all vertices and on $x_i$ as well as on all centroids $r_m$ calculated in 2.(b) (corresponding to $x_i^k$ at the start of the section), and is linearly interpolated in between. 

The absolute error of the integral on the cell $i$ is bounded from above by 
\begin{equation}\label{eq:leibnitz-absolut-error}
    |C_i|\,\sup_{x\in C_i}|f''(x)|\,\mathrm{diam}(C_i)^2
\end{equation}
due to Taylors formula and the relative error is bounded by 
\begin{equation}\label{eq:leibnitz-relative-error}
    \inf_{x\in C_i}|f(x)|^{-1}\,\sup_{x\in C_i}|f''(x)|\,\mathrm{diam}(C_i)^2\,.
\end{equation}

\subsubsection{The integration algorithm}
Using the data structure $\A$ we can formulate our integration algorithm.

The above observations all accumulate in the following result:

\begin{lem}
    Given a complete Voronoi diagram $\cM$ the algorithm \verb|IntegrateCell|($i,\cM,f$) yields for each cell $C_i$ the exact volume, as well as the exact area of all interfaces with its neighbors. Furthermore, it provides the exact integral of a function $F$ which is the linear interpolation of $f$ on the vertices, the cell centers $x_i$ and the midpoints $r_m$ calculated in Step 2.(b) of \verb|IterativeVolume|.
\end{lem}

\subsection{Heuristic Monte-Carlo integration}

We finally have a look at a modified Monte-Carlo integration that is able to save time significantly when it comes to the integration of functions.

In this case, we only need to compute the $d-1$ dimensional interface areas between $x_i$ and its neighboring cells $x_j$ first and infer from these areas the volume via 
$$\frac1d\sum_j \frac12|x_i-x_j|\;|\partial C_i\cap\partial C_j|\,.$$
We then take all $K_{ij}$ vertices $\cV_{ij}=\{\nu_1,\dots,\nu_{K_{ij}}\}$ belonging to the interface of $C_i$ and $C_j$ to calculate its centroid $y_{ij}=K_{ij}^{-1}\sum_k\nu_k$ like in the Leibnitz approach above. Furthermore, using the computed area $A_{ij}$ of the interface, we use the following modification of \eqref{eq:leibniz-int} to define an integral of $f(x)$ over $A_{ij}$
$$ \int_{A_{ij}}f(x)\dd x\approx\frac{d}{d+1}K_{ij}^{-1}\sum_k f(\nu_k)\;A_{ij}+\frac1{d+1}f(y_{ij})A_{ij}\,.$$
This mimics an integral over a linear interpolation of $f$ between the vertices of interface $i,j$ and the value at the centroid of $A_{i,j}$ in the same way as explained above. 
Applying once more \eqref{eq:leibniz-int} we can define an integral over the whole cell. 

The resulting integral is neither a Monte-Carlo intergral in the sense of Section \ref{sec:monte-int} nor the exact integral of a linear interpolant like in Section \ref{sec:leibnitz-int}. However, it does not require exhaustive calculations of determinants in high dimensions (like in Leibnitz), nor does it require that many function evaluations per cell like in Monte-Carlo. Instead, it combines the advantage of both approaches.  

\section{Computations and evaluation of performance}

In this section  we will carry out some numerical experiments demonstrating our techniques in this section.
We demonstrate the effectiveness of our incircle \raycast approach by contrasting it with the original bisection method.
We then show that our exhaustive search matches the performance of the state of the art \emph{qHull} solver.

Finally we compare the performance and accuracy of the proposed integration methods to each other and conclude with a discussion of their domains of application.

\label{sec:performance}

\subsection{Performance of the incircle \raycast}
\label{sec:perfbench}

We evaluate the performance of the previously suggested bisection search \cite{polianskii2020voronoi} against the newly proposed incircle \raycast (Section \ref{sec:raycast}).
It's important to keep in mind that the former search method is approximate and is dependent on the accuracy parameter $varepsilon$, which specifies the acceptable absolute error for the distance $t$ in refeq:t. As a result, in order to obtain results that are feasible for our experiment, we must decrease $\varepsilon$ for higher dimensions (because the distances increase).
Therefore, we compare the incircle \raycast search with and without the proposed initial point heuristic with the bisection search for two different tolerances, $\varepsilon = 10^{-4}$ or $\varepsilon = 10^{-8}$.
Since the nearest neighbour search consumes the majority of computation time, we evaluate their effectiveness by counting the number of nearest neighbour calls per vertex across various dimensions for $1000$ generators distributed uniformly in the unit cube:

\begin{table}[h]
    \centering
    \begin{tabular}{l|c|c|c|c}
method & $d=2$ & $d=3$ & $d=4$ & $d=5$\\ \hline
bisection ($10^{-4}$)& 8.15 & 6.91 & 6.74 & 6.58 \\
bisection ($10^{-8}$)& 8.18 & 7.24 & 7.25 & 7.28 \\
incircle& 2.70 & 2.82 & 2.76 & 2.76 \\
incircle heuristic& \textbf{2.41} & \textbf{2.46} & \textbf{2.54} & \textbf{2.62} \\
\end{tabular}
    \caption{Average nearest neighbour calls per vertex}
    \label{tab:nncalls}
\end{table}

We see that the incircle search with heuristic improves the number of nearest neighbor calls by a factor of $2.5 - 3.4$, decreasing with the dimensionality. The heuristic itself contributes an performance increase of about $10\%$.

We observe that the number of nearest neighbour calls is decreased by a factor of $3.4 - 2.5$ using the incircle search with heuristic, slowly decreasing with the dimension. The performance gain from the heuristic alone is roughly $10\%$.
It should be noted that if $\varepsilon$ is set too low, errors could accumulate to the point where "spurious vertices" are discovered, impeding the exhaustive search's ability to converge. The number of vertices found in the approximate and deterministic computations was consistently differing across the experiments indicating that even higher tolerances (and hence more nearest neighbour calls) would be necessary to obtain the qualitative exact Voronoi topology.

\begin{figure}
    \centering
    \includegraphics[width=0.6\textwidth]{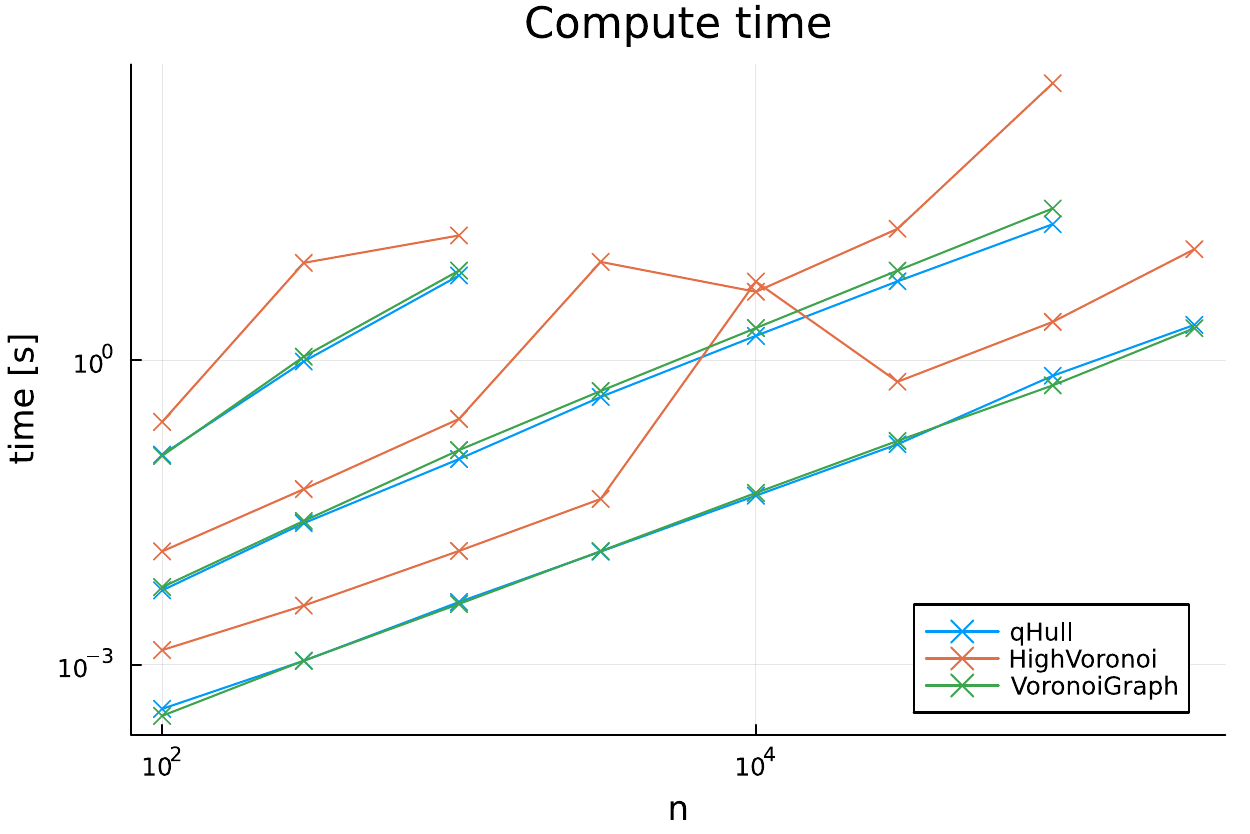}
    \caption{Comparison of the compute time of the Voronoi diagram in dimensions $d=2$ (lower three curves) and $d=6$ (upper three curves) over varying number of input nodes $n=(100,300,1\,000,...,300\,000)$.
    }
    \label{fig:times}
\end{figure}

We furthermore compute the raw compute times between the state of the art \emph{qHull} solver (using the delauney method) as well as both implementations of our algorithms. We do this by measuring the time to generate the voronoi diagram (resp. delauney triangulation) for different number of random generators $n=100, ..., 300\,000$ respectively $n=100,...,3000$ for dimensions $2$ and $6$.
In Figure \ref{fig:times} we see that our implementation (VoronoiGraph.jl) is matching \emph{qHull} in performance.
Whilst HighVoronoi is clearly slower it guarantees correct handling of degenerate tessellations and provides more flexibility in terms of boundary conditions etc. The spikes in its runtime are incurred by the special handling of possibly degenrater vertices.

To summarize, the incircle \raycast search not only provides an exact over an approximate solutions, but does this also at a fraction of the cost and with this new \raycast method the proposed exhaustive search matches the performance of established voronoi compututation algorithms.

\subsection{Performance of the integration routines}
We will now evaluate the performance and quality of our suggested integration algorithms. 
For simplicity of presentation we write \textbf{MC}  for Monte-Carlo integration, \textbf{P} for Leibnitz integration (P stands for Polygon or Polytope) and \textbf{HMC} for heuristic Monte-Carlo integration.

We start with comparing the time needed computing integrals using the respective methods and continue with a discussion of the approximation quality.

We used \verb|HighVoronoi.jl| for the calculations, as this package supports the restriction of volume calculation to a unit cube and hence all values for volumes, interfaces or integrals in our simulation are finite. 
When we write about \verb|somefile.jl| about we always refer to a file with the corresponding name in that repository. 

\subsubsection{Computation time}

\begin{figure}
    \centering
    \includegraphics[width=0.32\textwidth]{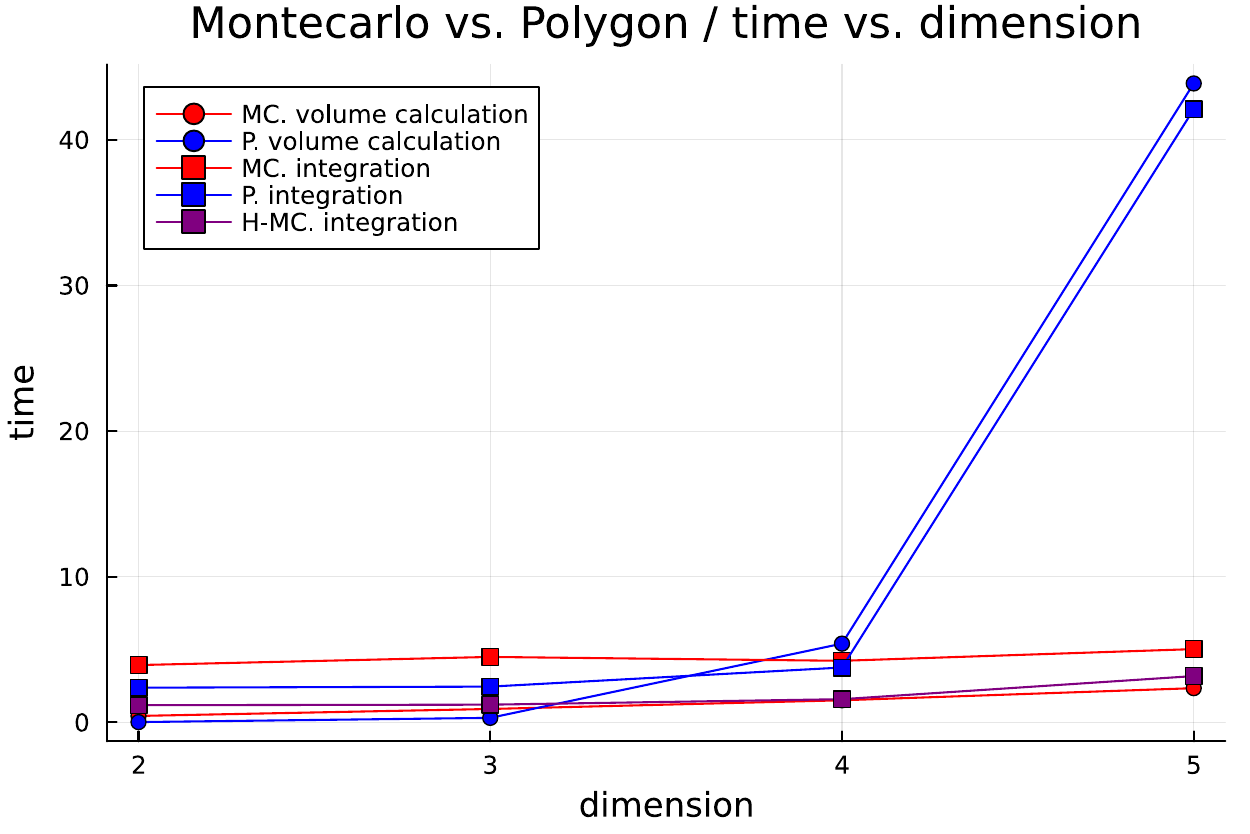}
    \includegraphics[width=0.32\textwidth]{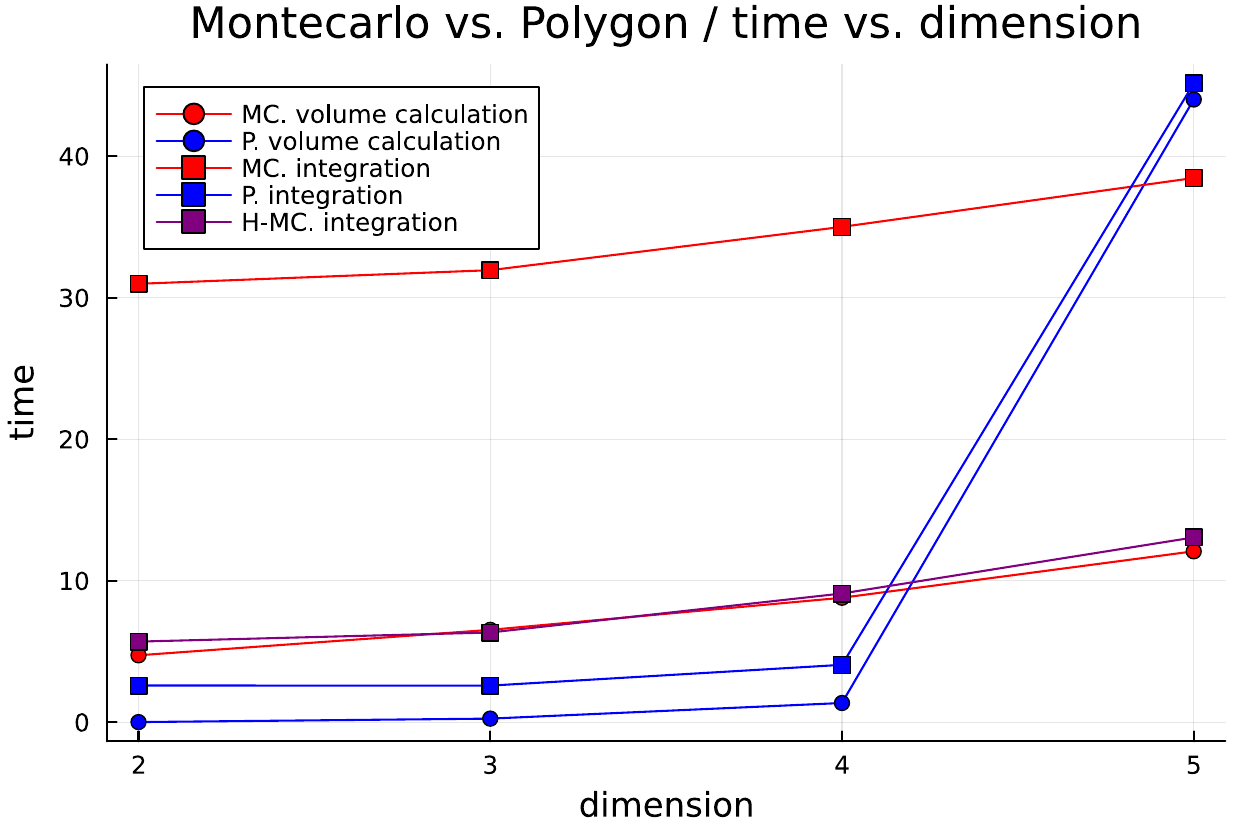}
    \includegraphics[width=0.32\textwidth]{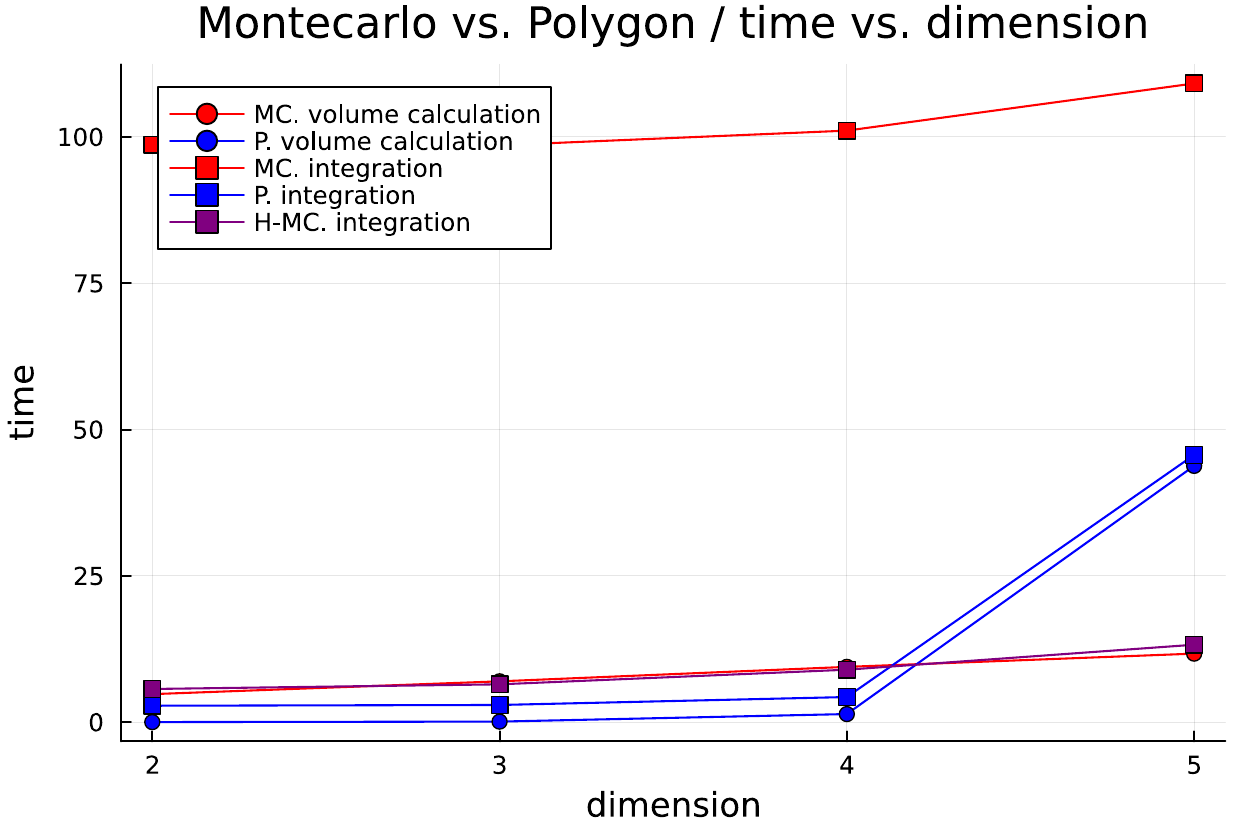}
    \caption{Compute time of the volume and integral calculations using the Polygon (\textbf{P}),  Monte-Carlo ( \textbf{MC} ) and heuristic MC (\textbf{HMC}) methods. 
    The number of samples (\#rays, \#t) for  \textbf{MC}  are, from left to right: (1\,000, 2), (10\,000, 2) and (10\,000, 10). }
    \label{fig:performance-all-2-5}
\end{figure}


In \verb|test-integration.jl| we distribute 1000 nodes iid in the unit cube for each dimension $d=2$ to $d=5$ and calculates the corresponding Voronoi diagrams.
As integrand we choose the function $f(x)=sin(x_1^2)$, which is a rather smooth function yet bearing some some computational cost. 
\begin{itemize}
    \item the Monte-Carlo (\textbf{MC} ) and Polygonal (\textbf{P}) volumes and areas
    \item as well as the Monte-Carlo, Polygonal and heuristic Monte-Carlo (\textbf{HMC}) volume and area integrals for $f(x)$.
\end{itemize}

Since the cost of \textbf{MC}  strongly depends on the number of rays, $n$, and (for the integration of a function, eq. \eqref{eq:full_MC_integration}) the number of volume subsamples, $m$, we performed simulations for $n=10^3$ and $n=10^4$ rays as well as for $m=2$ and $m=10$. The number of rays was also applied to \textbf{HMC} for a meaninful comparison, but we note that the choice of $m$ has no effect on HMC. We restricted the analysis to $d\le5$ since in $d=6$ the Leibnitz computation takes already more than 1 hour.

As we can see in Figure \ref{fig:performance-all-2-5} the costs of \textbf{P} steeply increase from 4 to 5 dimensions. This is no surprise as the number of edges, which get subdivided to derive the simplicial complex,increases superexponentially with $d$.

As to be expected, the cost of the Monte-Carlo integration scales with their parameters $n$ and $m$. 
The \textbf{HMC} function integration builds upon the Monte-Carlo area computation. We see that the \textbf{HMC} integration takes about the same time as the \textbf{MC} area/volume computation, indicating that the additional cost for evaluating $f$ at the vertices is neglible.
However, when looking at \textbf{MC} \emph{volume} we observe a clear increase in cost, which is explained by the huge amount of additional function evaluations $n\times m$.




\subsubsection{Approximation quality of volumes and volume integrals}

\begin{figure}
    \centering
    \includegraphics[width=0.32\textwidth]{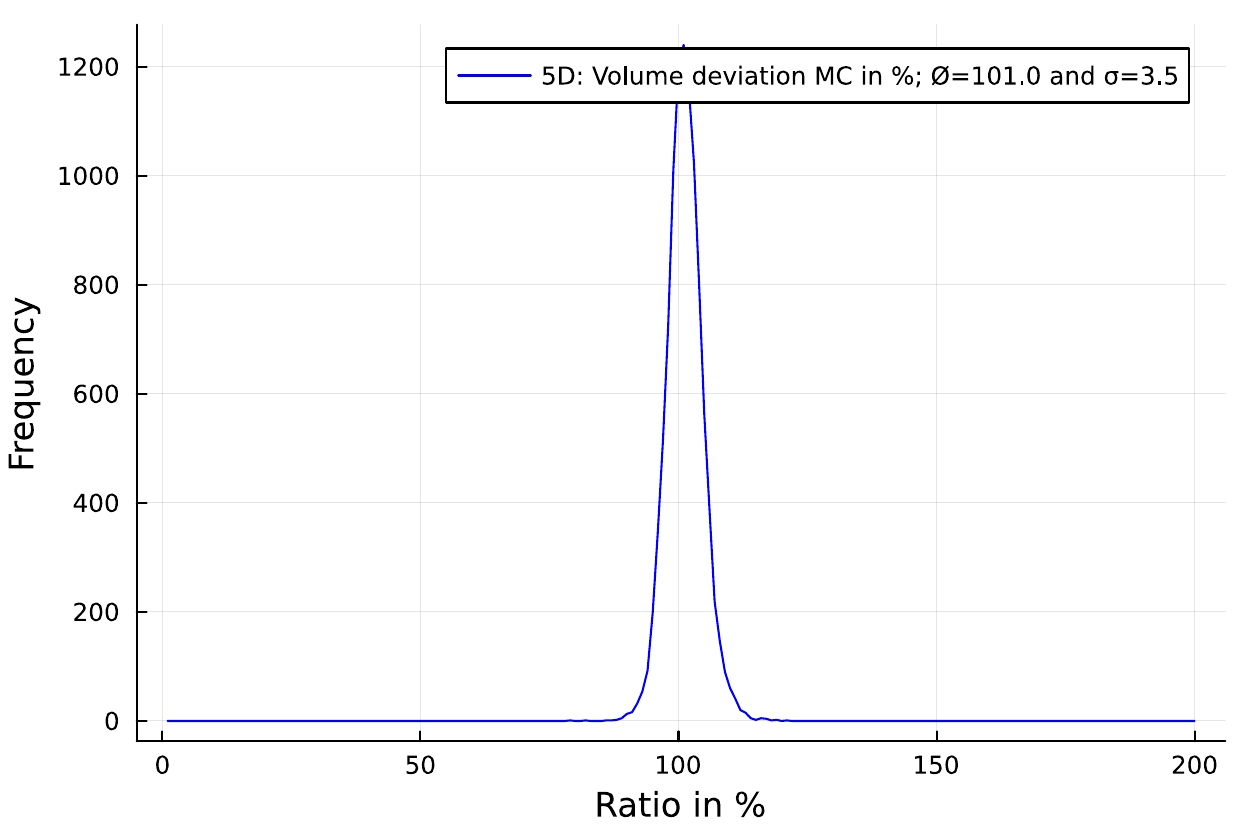}
    \includegraphics[width=0.32\textwidth]{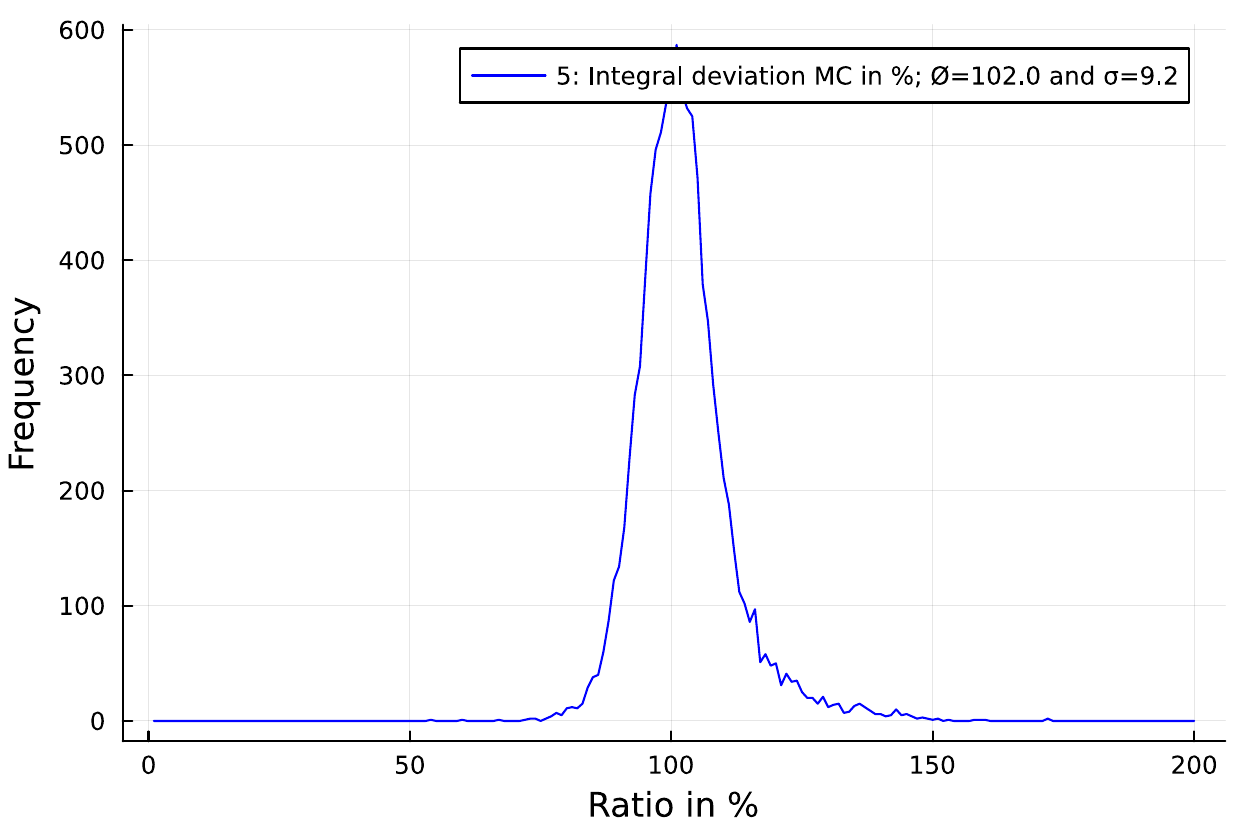}
    \includegraphics[width=0.32\textwidth]{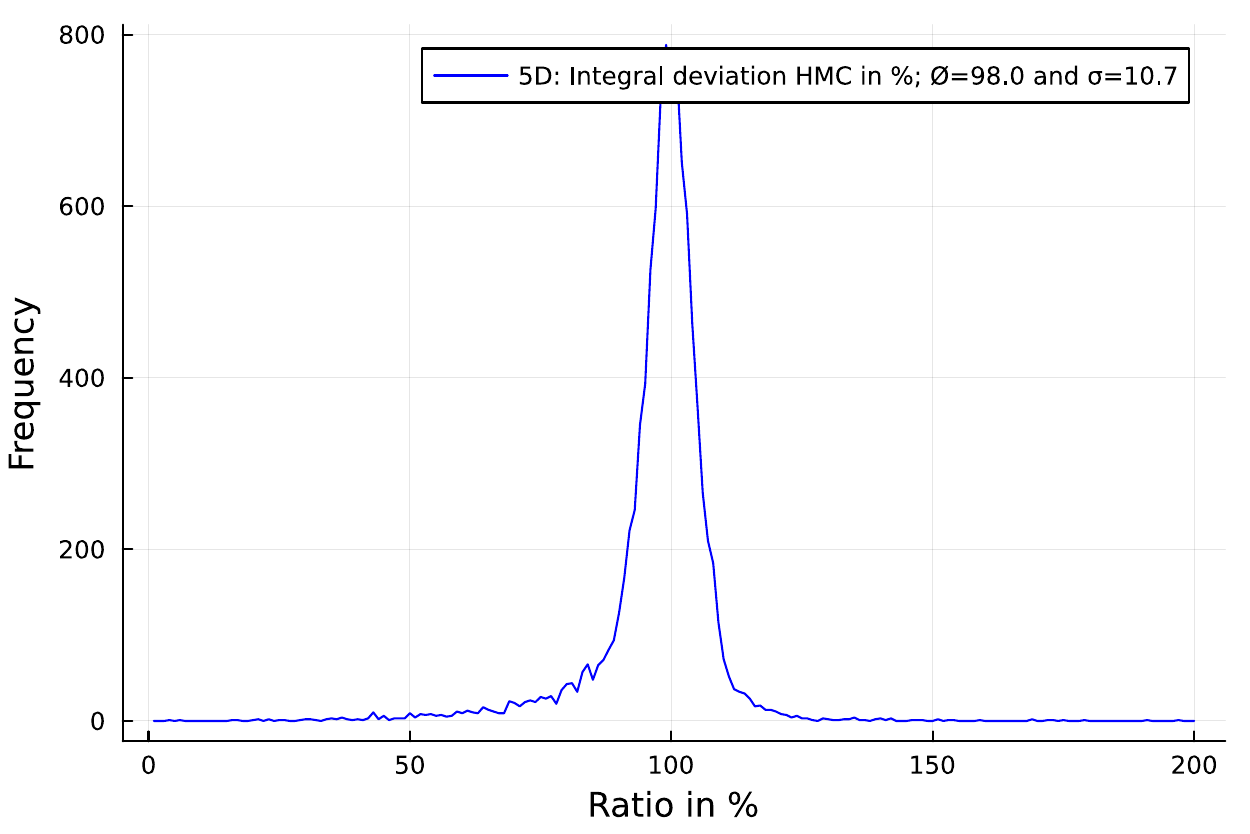}
    \caption{Deviation of \textbf{MC} volume from exact volume (left), deviation of \textbf{MC} integral from \textbf{P} integral (middle) and deviation of \textbf{HMC} integral from \textbf{P} integral (right). All plots are for $n=1000$ rays in the \textbf{MC} method and $m=2$ sample points along each ray. }
    \label{fig:vol-deviations}
\end{figure}

Note that the Polygonal method \textbf{P} computes the exact volumes of the cells. To the left of Figure \ref{fig:vol-deviations} we compare the relative error of the volumes approximated with \textbf{MC} in $d=5$ dimensions for $N=10\,000$ nodes.

Since we have no access to the the true volume integrals, we can only compare the three approximative integration methods to each other.
We expect \textbf{P} to be more accurate then \textbf{HMC} in general, since both integrate linear approximations but \textbf{P} uses more interpolation points and exact instead of approximate volumes.
We therefore provide the histograms of the relative deviations of the \textbf{P} vs. \textbf{MC}  and \textbf{P} vs. \textbf{HMC} to the right of Figure \ref{fig:vol-deviations}.

It appears that the \textbf{MC} integrals systematically overestimates while the \emph{HMC} integral underestimates relative to \textbf{P} integration method by around 2\%. It is not clear how this happens but it might be due to the shape of the function $sin(x^2)$. 
We also observe that the over- or underestimation goes in hand with a tail of the distribution on the respective side.


\subsubsection{Approximation quality of the surface area}

\begin{figure}
    \centering
    2D) \includegraphics[width=0.32\textwidth]{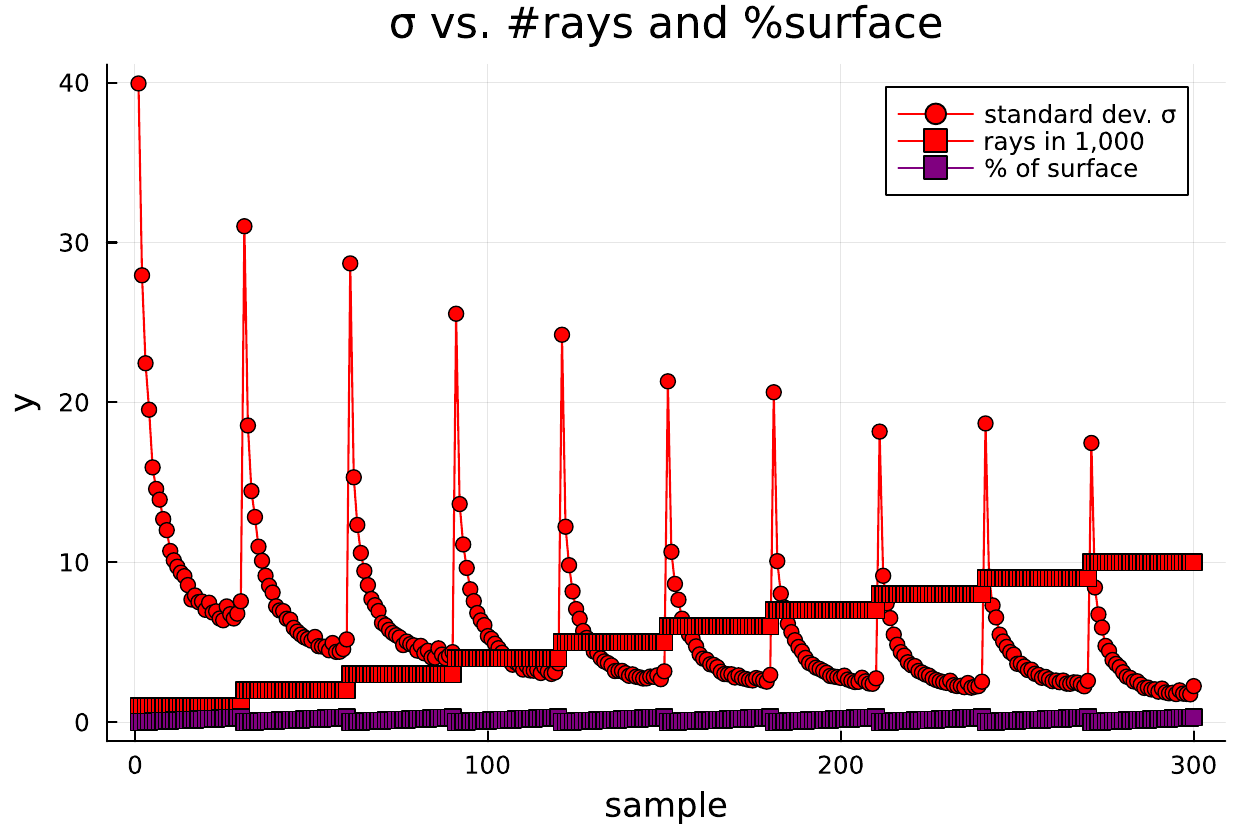} $\qquad$
    3D) \includegraphics[width=0.32\textwidth]{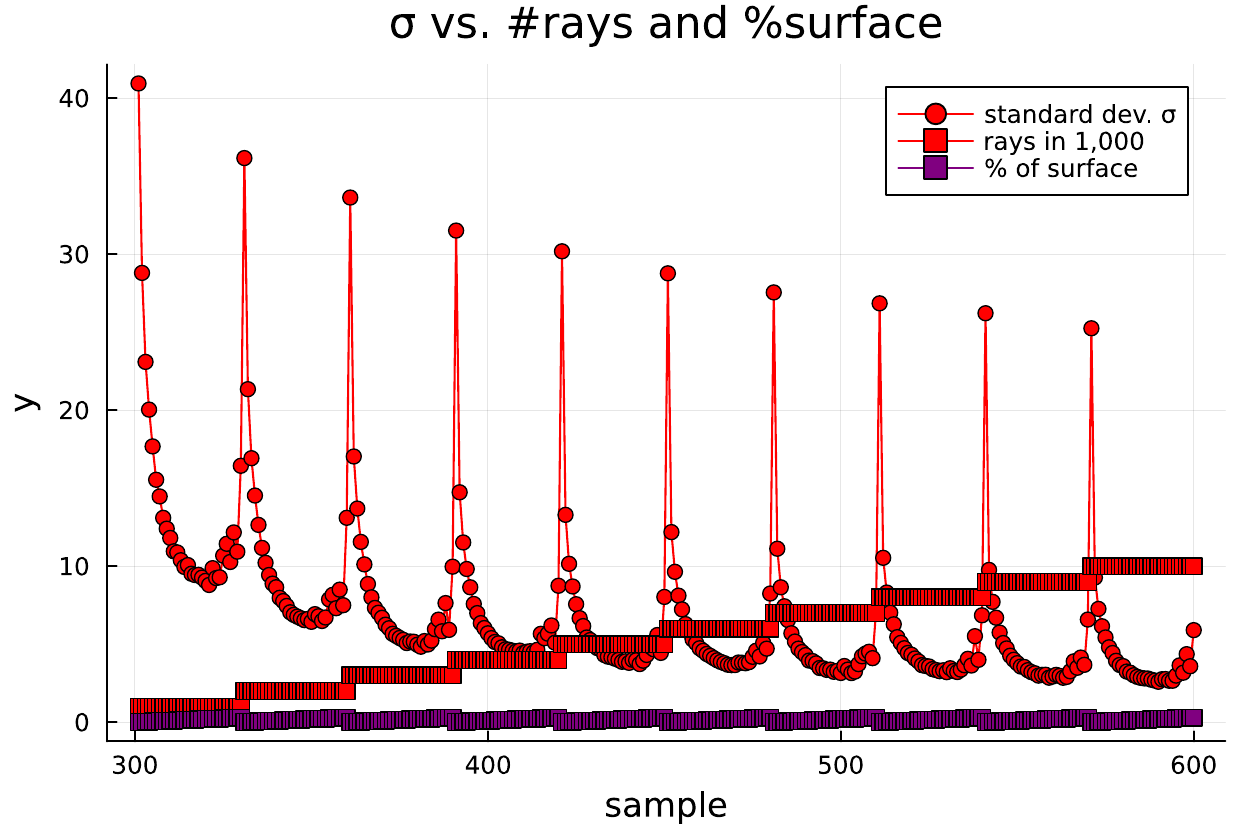}\\
    4D) \includegraphics[width=0.32\textwidth]{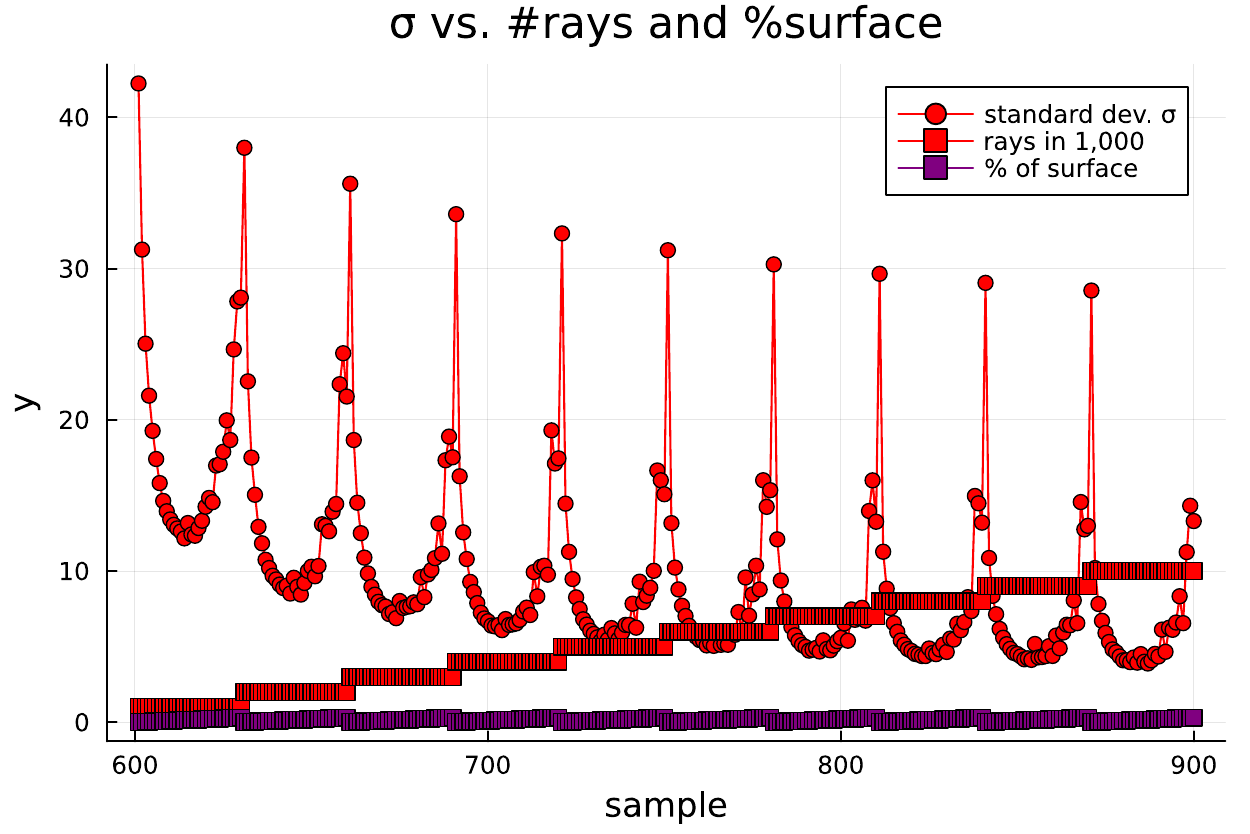}$\qquad$
    5D) \includegraphics[width=0.32\textwidth]{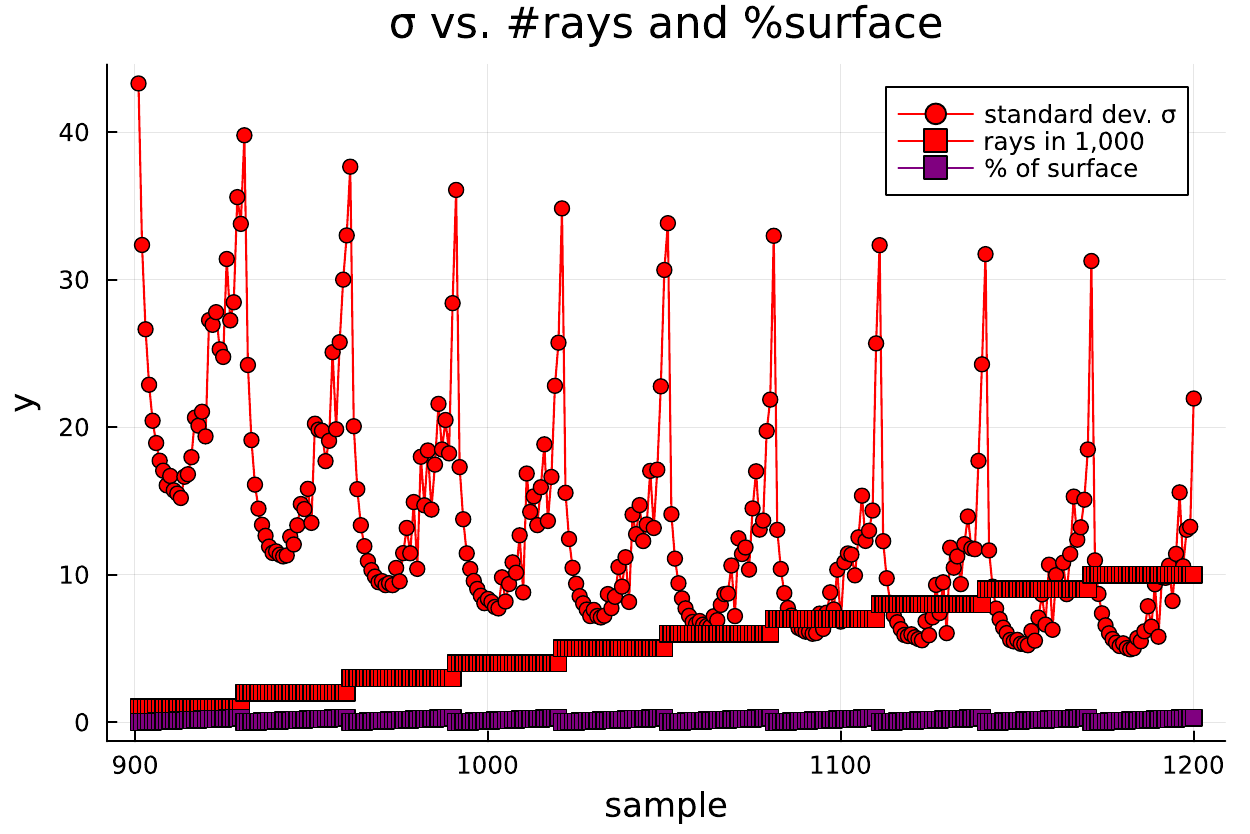}
    \caption{Each of the four figures shows the mean deviation of Monte-Carlo area vs. exact area in 2D to 5D. Each ,,bow'' corresponds to data for resp. 1\,000, 2\,000, ... 10\,000 rays per cell. Inside each bow the relative deviation is shown for the faces of relative contribution of resp. 1\% to 30\% of the total cell surface. 
    }
    \label{fig:all-data-area-var}
\end{figure}

The expected error for individual interface integrals, i.e. over the surfaces between two cells, is expected to be much higher than the volume volume integrals themselves, given that the rays used to compute a volume integral distribute their contribution over their individual interfaces.
For example, when computing a full surface integral with  $10\,000$ rays (samples) the the integral over a interface that covers only 1\% of the total surface relies on merely 100 rays. Correspondingly, in this particular case the variance can be expected to be 100 times larger (factor 10 in the standard deviation) according to the law of large numbers.

As mentioned above, the \textbf{P} method is exact concerning volume and area (besides machine precision). When it comes to integral computation, it is as exact as a piecewise affine approximation of the original function, i.e. it can be expected to be as precise as $\frac13f''(x)\mathrm{diam}(C_i)^3$ on cell $C_i$.  

In the following we compare the variance of the relative errors for the area estimation.
The data is computed over 4 different realizations of Voronoi grids for $1\,000$ random nodes in dimensions $d=2,\dots, 5$ and differing number of Monte-Carlo rays, $1\,000, 2\,000, \dots, 10\,000$.
We furthermore grouped the faces into $30$ bins according to their 
relative contribution to the total cells surface, 
$I:=\frac{|\sigma_{ij}|}{|\partial C_i|}$, for 
$1\% \le I \le 30\%$ 
\footnote{See the file  test-integration-area-accuracy.jl}.

The results are shown in Figure \ref{fig:all-data-area-var}.
We can observe the expected square-root depende of the noise on the number of Monte-Carlo rays. Regardless of the dimension or number of rays in \textbf{MC}, the deviation of the \textbf{MC} area from the \textbf{P} area is large for small area fractions, decreases to a minimum with increasing area fraction and from dimension $d\geq3$ rises again for large area fractions. 
The minimum of this curve seems to be close to $\frac1{2d}$ which indicates that the Monte-Carlo estimator works best for areas close to a face of a cube.  

The explanation we have is that small area fractions are hard to be measured precisely by a random algorithm as the probability to have a representative sample is low. On the other hand, high area fraction means that the area is expanding into the ,,outskirt'' of the respective piece of surface, the measure becomes unprecise due to the flat angle at which a ray intersects the area. 

Note however that, depending on the application, the uncertainty of the individual faces area or even their function integrals may average out over patches of neighbouring cells.

\subsubsection{Approximation quality of surface integrals}

In order to compare \textbf{MC}, \textbf{HMC} and \textbf{P} integrals over interfaces we follow a similar strategy as for the areas. 
Using the code in \verb|test-integration-mc-vs-hmc.jl| we collect data for $1\,000$ nodes in dimensions $1$--$5$ for two mutual methods on the same Voronoi grid. We then compare the integral values $I_1$ and $I_2$ of both methods on each interface and determine the deviation as $1+2\frac{I_1-I_2}{I_1+I_2}$.

The quality of interface integrals of \textbf{MC} and \textbf{HMC} methods decreases with the dimension due to the increase of interfaces per cell according to Table \ref{table:empscaling} above, resulting in less rays per interface and decreasing resolution. Hence in our presentation and analysis we focus on $d=5$.

\begin{figure}
    \centering
    \includegraphics[width=0.32\textwidth]{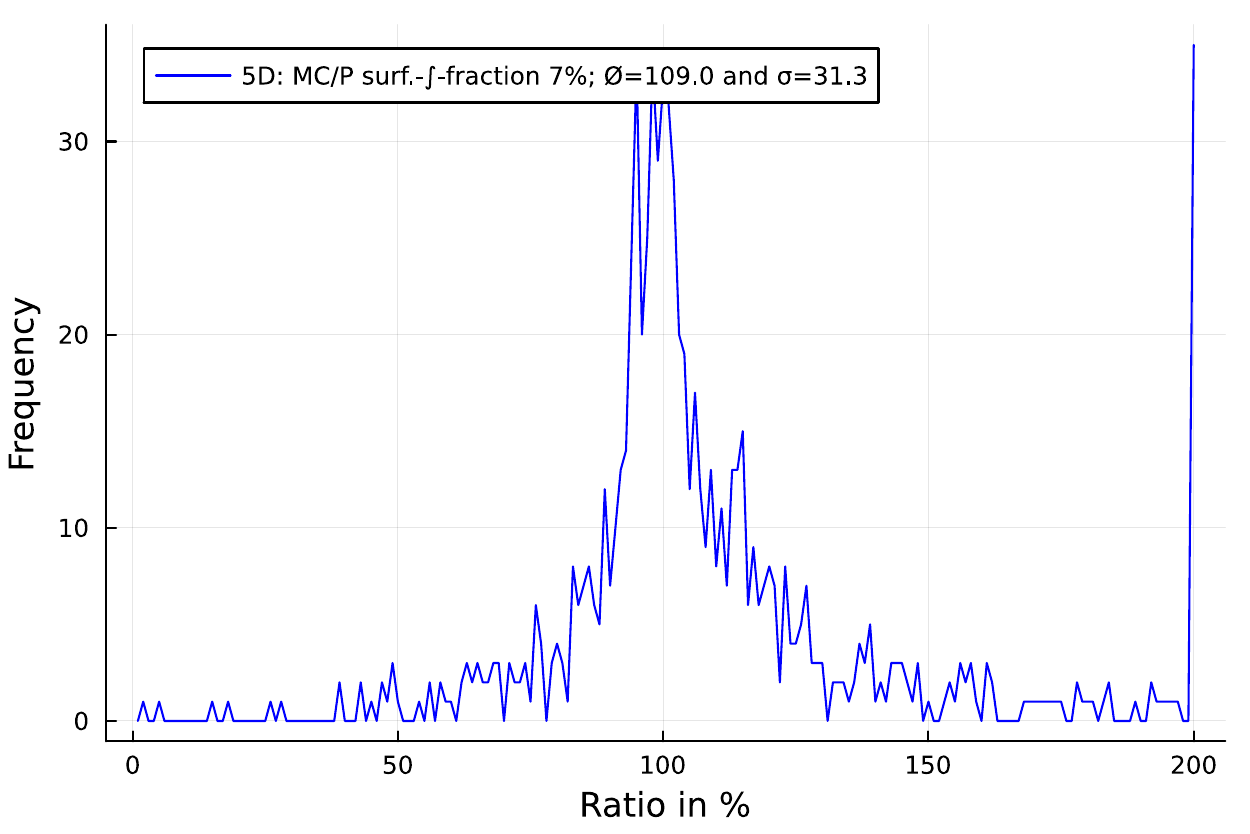}
    \includegraphics[width=0.32\textwidth]{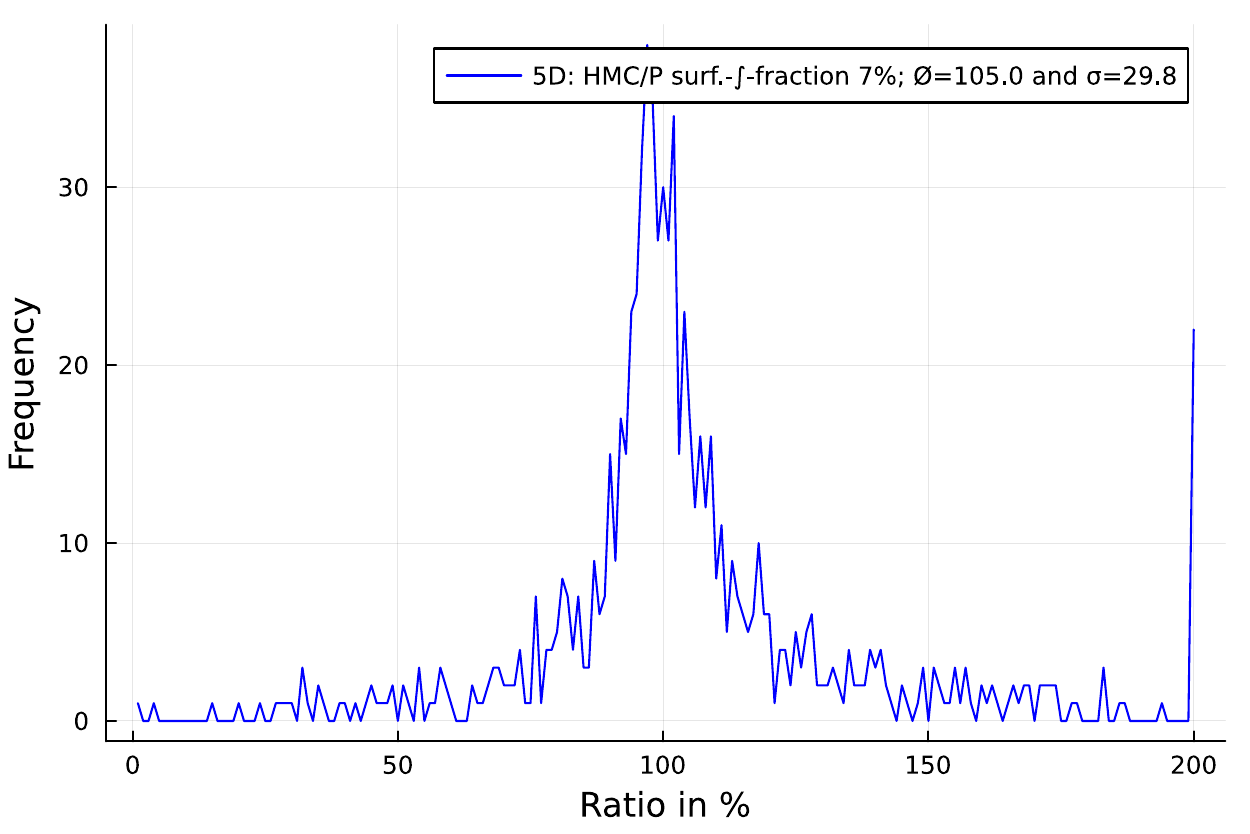}
    \includegraphics[width=0.32\textwidth]{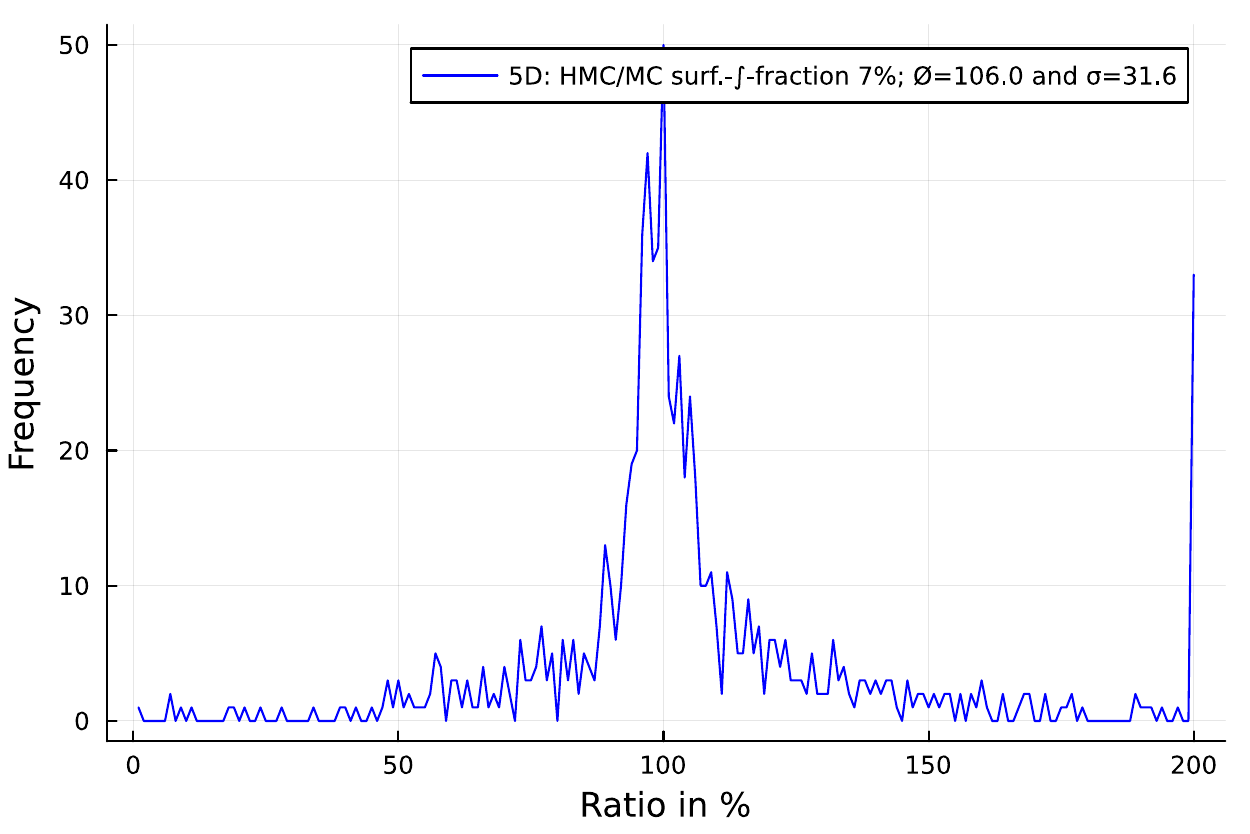}
    \caption{Comparison of integration methods for areas of approximately 7\% of total cell-surface area. 
    }
    \label{fig:compare-dim-7}
\end{figure}
In Figure \ref{fig:compare-dim-7} we provide the graph for an surface integral fraction of 7\% in $d=5$ for \textbf{MC} vs. \textbf{HMC}, \textbf{MC} vs. \textbf{P} and \textbf{HMC} vs. \textbf{P}. We observe that the profiles do not differ significantly. Hence we can conclude that they more or less provide the same quality of approximation of the true integral value. 

\begin{figure}
    \centering
    \includegraphics[width=0.32\textwidth]{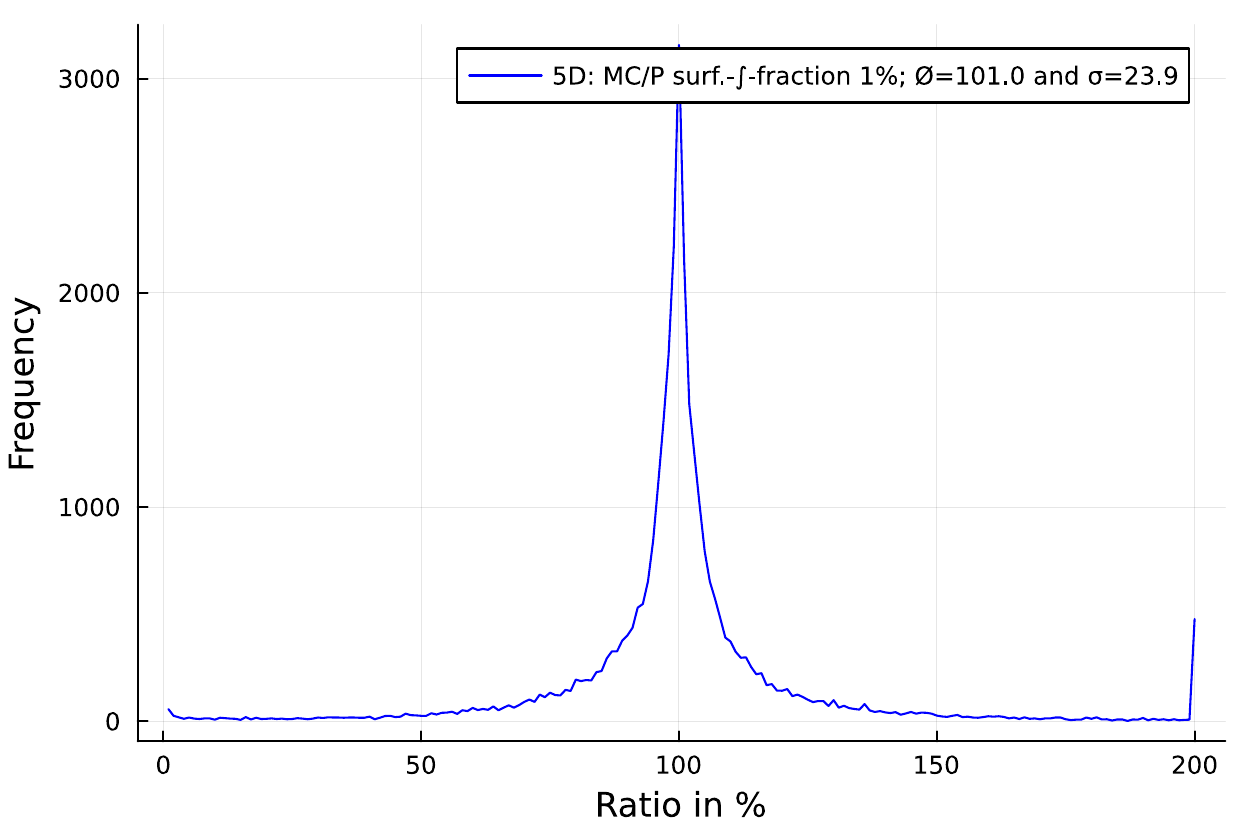}
    \includegraphics[width=0.32\textwidth]{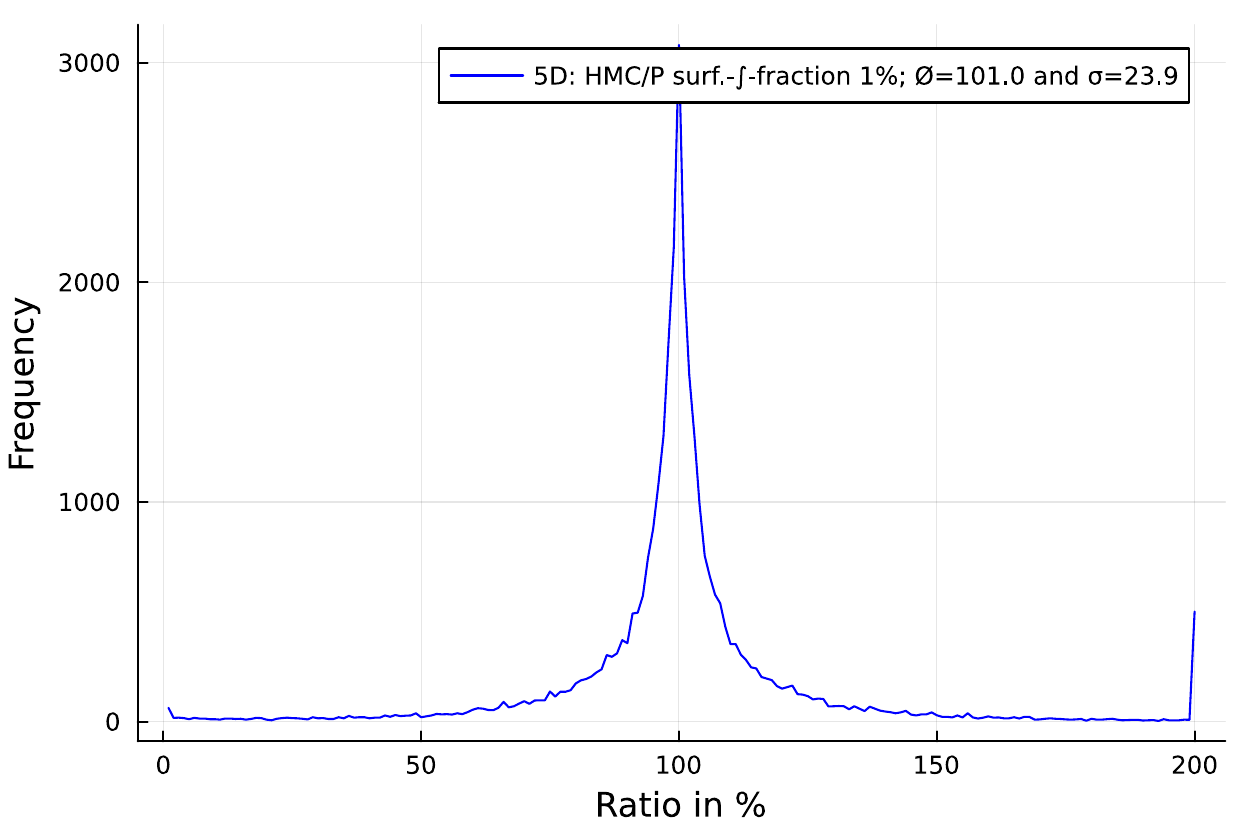}
    \includegraphics[width=0.32\textwidth]{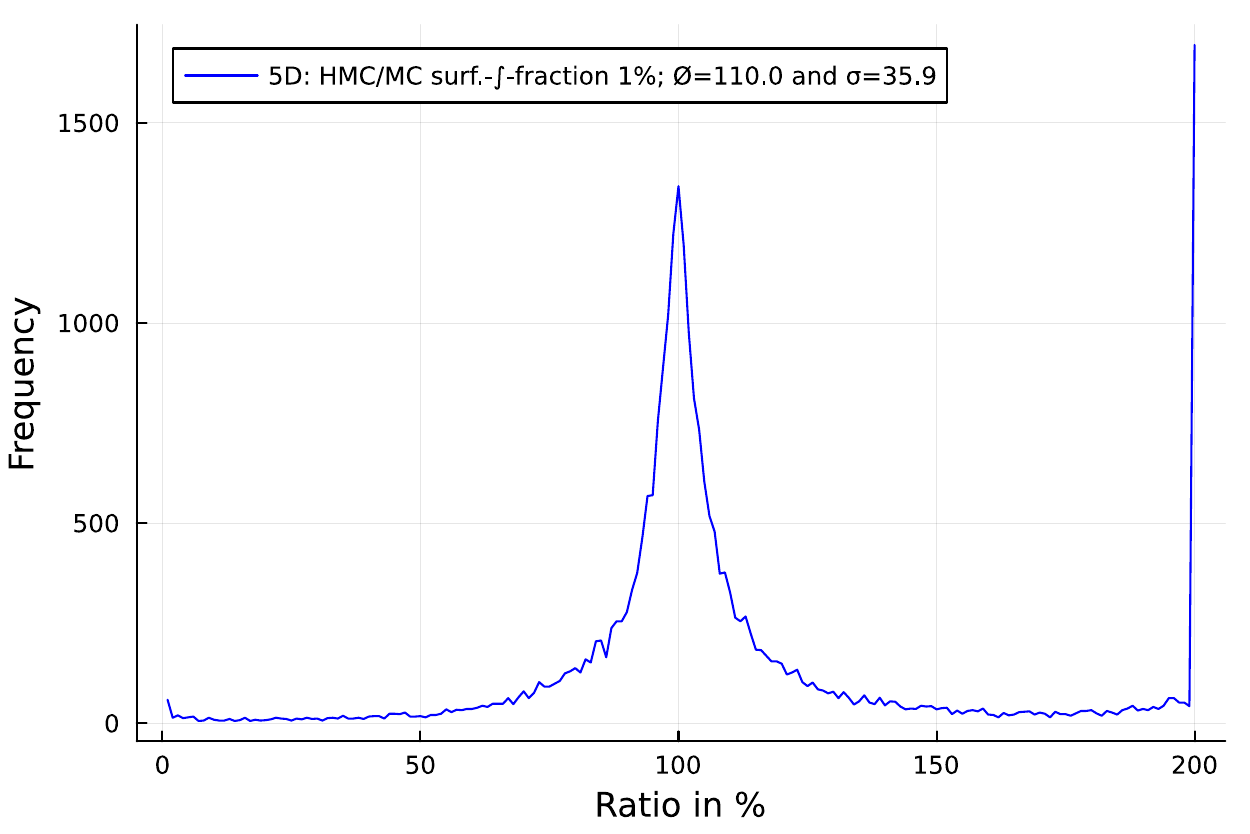}
    \caption{Comparison of integration methods for areas of approximately 1\% of total cell-surface area. 
    }
    \label{fig:compare-dim-1}
\end{figure}

For smaller interface integral fraction, we however observe a difference. In Figure \ref{fig:compare-dim-1} we provide the same comparison for a percentage of 1\% or less and find that \textbf{HMC} and \textbf{MC} compare the same way to \textbf{P} but the difference between \textbf{HMC} and \textbf{MC} as significantly worse mean deviation $\sigma$.

We note at this place that the \verb|HighVoronoi| package makes it possible to take a \textbf{MC} integration and calculate \textbf{HMC} data for exact the same volume and interface data. 

\begin{figure}
    \centering
    \includegraphics[width=0.4\textwidth]{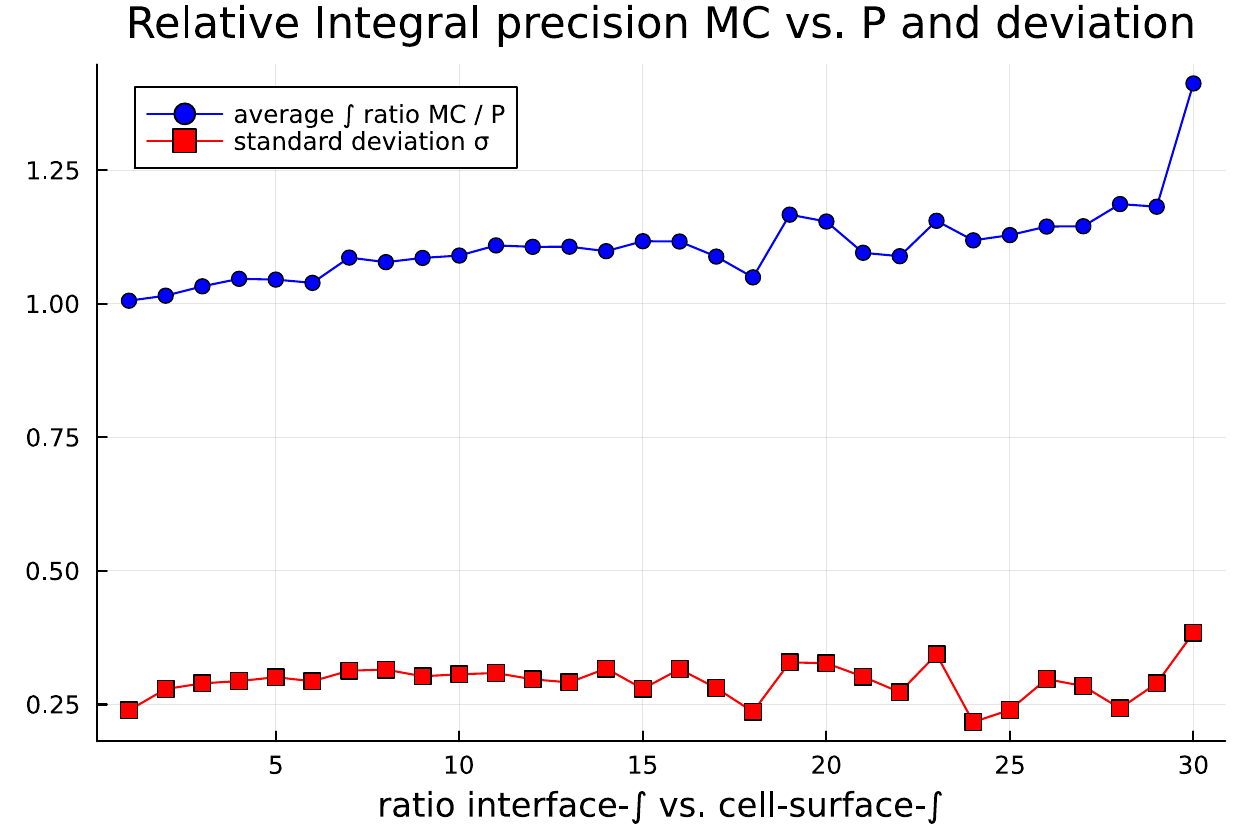}
    \caption{The variation of \textbf{MC} integral vs. \textbf{P} integral in 5 dimensions for different relative area fractions. It turns out that the \textbf{MC} integral is systematically higher. The standard deviation is approximately constant but starts to oscillate for high  relative area fractions, i.e. for a percentage of more than 16\%. This can probably be associated with a small number of samples for large relative sizes, since the surface of a cell is scattered into approximately 90 parts in $d=5$.}
    \label{fig:general}
\end{figure}

When it comes to higher percentage of integral values per interface, we point out that Figure \ref{fig:general} indicates that the low number of samples in this case makes data unreliable.

\subsection{Comparison of the suggested methods}
Of the three suggested integration methods, Monte-Carlo (\textbf{MC}), Polygonal (\textbf{P}) and heuristic Monte-Carlo  (\textbf{HMC}), only \textbf{P} is able to compute the areas and volumes exact. It furthermore allows to exactly integrate linear interpolations to a function $f$. The interpolation nodes are the vertices of the simplicial complex generated by the Voronoi, i.e. even finer then the Voronoi Diagram itself and the error scales with the cell size via $\frac13f''(x)\mathrm{diam}(C_i)^3$. This allows for good accuracy but scales extremely bad with the dimension.

On the other hand we have \textbf{MC}, which itself scales independently of the dimension but does not use any additional structure of the function $f$ and therefore is expected to provide less accurate estimations for the same number of $f$ evaluations. However, it is known that the error scales with the number of samples $N$ via $\frac{1}{\sqrt{N}}$.

Finally there is the \textbf{HMC} which estimates the areas/volume via Monte-Carlo but averages over the function evaluations of the vertices, mimicking  the averaging of \textbf{P} over the simplices. We expect this to be more efficient then \textbf{MC} and \textbf{P} for medium dimensions, where \textbf{P} is unfeasable but the number of vertices is still low enough to benefit from the interpolation-like aspects. However we have to admit that the averaging is merely heuristically motivated and we don't know of any precise error estimates.

We summarize this in the following table

\begin{table}[h]
    \centering
    \begin{tabular}{c|c|c|c}
         Method & Accuracy Integrals & Accuracy Volumes & Scaling with $d$ \\
         \hline
         \textbf{P} & very good &exact & very bad \\
         \textbf{HMC} & good (?) & using \textbf{MC} & bad \\
         \textbf{MC} & $\sqrt{N^{-1}}$ & $\sqrt{N^{-1}}$ & good  
    \end{tabular}
    \caption{Comparison of the integration methods}
    \label{tab:my_label}
\end{table}

\section{Conclusion}

The fundamental contribution of this paper is the new incircle \raycast algorithm (Algorithm \ref{alg:raycast}).
It improves on the original bisection \raycast \cite{polianskii2020voronoi} not only by providing the exact position of the vertices, but also reduces the required number of nearest neighbour calls that make up the major cost of the computation by a factor of $~3$.
Building on this we furthermore introduce a new exhaustive search that allows the computation of the full and exact Voronoi diagram in medium dimensions, matching or even exceeding the performance of the state of the art algorithm qHull.

We furthermore introduced three different integration methods. The most general is a pure Monte-Carlo method using the \raycast to approximate surface and volume integrals even in very high dimensions. Then we have a polygonal Leibnitz rule based method that subdividies the cells into simplices and gives exact results for areas, volumes and linear interpolants but is limited to low dimensions. We combine these approaches into a heuristic Monte-Carlo which allows to integrate by function evaluation on the vertices and Monte-Carlo area/volume approximation suitable in medium dimensions.

\section{Acknowledgements}
This research has been funded by Deutsche Forschungsgemeinschaft (DFG) through grant CRC 1114 "Scaling Cascades in Complex Systems", Project Number 235221301, Projects B05 "Origin of scaling cascades in protein dynamics" and C05 "Effective models for interfaces with many scales".

\bibliography{sample}

\newpage
\section*{Appendix: Algorithms}

\begin{algorithm} 
\caption{Incircle \raycast}\label{alg:raycast}
\begin{algorithmic}[1]
\Require 
$\eta$ indices of the known generators for the vertex \\
$r$ a reference point equidistant to all generators $\eta$ \\
$u$ the search direction, orthogonal to the subspace spanned by $\eta$ \\
$X$ the set of all generators
\Ensure $(\sigma',r')$ are generators of the vertex in direction $u$
\Function{\raycast}{$\eta, r, u, X$}

\State $r \gets \Call{InitialHeuristic}{r, u, X_{\eta_1}, \eta}$
\State $\Call{skip}{i} = \left<X_i, u\right> \le \max_{g\in\eta} \left<x_g, u\right>$
\Comment{restrict search to the correct halfspace}
\State $i \gets \Call {nn}{r, \Call{skip}{}}$ 
\Comment{find initial generator candidate}
\Loop

\State $x \gets X_i$
\State $t \gets \frac{|| r-x||^2 - || r-x_0||^2}{2 \left<u, x-x0\right>}$ 
\State $r' \gets r + t\cdot u$
\Comment{compute candidate vertex}
\State $j \gets \Call {nn}{r', \Call{skip}{}} $
\If{$i=j$} \Comment{check for incircle criterion}
\State $\sigma' \gets \eta \cup i$
\State \Return $(\sigma', r)$
\ElsIf{\Call{nn}{} did not return a neighbour}
    \State \Return nothing
\Else
\State $i\gets j$ \Comment{repeat search from improved candidate}
\EndIf
\EndLoop
\EndFunction
\\
\Function{nn}{$x$, \Call{skip}{}}
    \State \Return the index $i$ to the nearest neighbor of $x$ among $X$, such that the predicate \Call{skip}{$i$} is false.
\EndFunction
\\
\Function{InitialHeuristic}{$\candidate, u, \node, \eta$}
    \State $d \gets \text{length}(\eta)$
    \State $\candidate \gets \candidate + u \left<u, \node - \candidate\right>$
    \Comment{projection onto the face}
    
    \State $\candidate \gets \candidate + u \frac{|\candidate - \node|}{\sqrt{(d-1)(d+1)}}$
    \Comment{shift to simplex center using rescaled radius}
    \State \textbf{return} $\candidate$
\EndFunction
\end{algorithmic}
\end{algorithm}

\begin{algorithm}
\caption{VoronoiGraph - compute the voronoi diagram from a cloud of points}\label{alg:explore}
\begin{algorithmic}[1]
\Function{VoronoiGraph}{$X$}
    \State $V \gets \emptyset$ 
        \Comment{\makebox[\commentlen][l]{set of vertices}}
    \State $B \gets \emptyset$ 
        \Comment{\makebox[\commentlen][l]{set of boundary rays}}
    \State $\EC \gets \emptyset$ 
        \Comment{\makebox[\commentlen][l]{dictionary storing the edge visited count}}
    \State $Q \gets \{\Call{Descent}{\{\,\},x_1,X}\}$ 
        \Comment{\makebox[\commentlen][l]{intialize queue with a vertex}}
    \\
    \While{$Q \text{ is not empty}$}  
            \Comment{\makebox[\commentlen][l]{loop over all discovered vertices}}
      \State $(\sigma, r) \gets \Call{pop}{Q}$ 
        \Comment{\makebox[\commentlen][l]{take and remove an element from queue}}
      \State $V \gets V \cup \{(\sigma, r)\}$ 
        \Comment{\makebox[\commentlen][l]{and add it to known vertices}}
      \For{$\eta \in E(\sigma)$ where $\EC[\eta] < 2$} 
            \Comment{\makebox[\commentlen][l]{go over all unexplored edges}}
        \State $d \gets \Call{SearchDirection}{\sigma, \eta}$
        \State $( \sigma ', r') \gets \Call{\raycast}{\eta, r, d, X}$ 
            \Comment{\makebox[\commentlen][l]{compute the vertex}}
        \If{$(\sigma', r')$ is a vertex}
            \State $Q \gets Q \cup \{(\sigma', r')\}$ 
                \Comment{\makebox[\commentlen][l]{add discovered vertex to queue}}
            \For{$\eta' \in E(\sigma')$} 
            \State $\EC[\eta'] \gets (\EC[\eta'] \lor 0) + 1$ 
                \Comment{\makebox[\commentlen][l]{create and/or increase edge counter}}
        \EndFor
        \Else
            \State $B \gets B \cup \{(\sigma, d)\}$
                \Comment{\makebox[\commentlen][l]{add ray to boundary}}
        \EndIf
    \EndFor
   \EndWhile
   \\
   \State \textbf{return} $V, B$
\EndFunction
\\
\Function{E}{$\sigma$}
    \State \textbf{return} all subsets of $\sigma$ omitting exactly one element
\EndFunction
\\
\Function{SearchDirection}{$\sigma, \eta$}
    \State \textbf{return} a unit vector $d$ orthogonal to the plane spanned by $X_\eta$ and pointing away from the additional generator $X_{\sigma \setminus \eta}$
\EndFunction
\\
\Function{Descent}{$\sigma, r, X$}
    \State $u \gets \text {uniform sample from } S^{d-1}$ 
        \Comment{e.g. normalized multivariate normal}
    \State $u \gets u - \text{ orthogonal projection of u onto the plane spanned by }X_\sigma$
    \State $(\sigma, r) \gets \Call{\raycast}{\sigma, r, u, X)}$ 
    \If{length$(\sigma) = d+1$}
        \State $\Return (\sigma, r)$
    \Else
        \State \Return $\Call{Descent}{\sigma, r, X}$
    \EndIf
\EndFunction
\end{algorithmic}
\end{algorithm}

\begin{algorithm}[h]
\caption{Monte-Carlo integration over a cell's surface and volume}
\label{alg:montecarlo}
\begin{algorithmic}[1]

\newcommand{\pluseq}{\mathbin{{+}{=}}}

\Function{MonteCarlo}{$i,X,f,n,m$}
\State $V, F_V \gets 0$
\State $A, F \gets $ Dictionary mapping each neighbor index to 0
\For{$k=1$ to $n$}
    \State $u \gets \text{uniform sample from }S^{d-1}$
    \State $(\sigma, r) = \Call{\raycast}{(i), X_i, y, X}$ 
    \State $l \gets |r - X_i|$
    \State $j \gets \sigma \setminus (i)$
    \State $n \gets \frac{X_i - X_j}{|X_i - X_j}$
    \State $A[j] \pluseq \frac{l^{d-1}}{\left<n,u \right>} $
    \State $V \pluseq l^d$
    \State $F[j] \pluseq f(r) \frac{l^{d-1}}{\left<n,u \right>}$
    \For{$l=1$ to $m$}
        \State $t \gets \Call{rand}{(0,1)}$
        \State $F_V  \pluseq f((1-t)X_i + t r)) t^{d-1}l^d\cdot $
    \EndFor
\EndFor
    \State $A \gets A \frac{S_{d-1}}{n}$
    \State $F_\partial \gets F_\partial \frac{S_{d-1}}{n}$
    \State $V \gets V \frac{S_{d-1}}{dn}$ 
    \State $F_V \gets F_V \frac{S_{d-1}}{nm}$
    \State \Return $A, V, F_\partial, F_V$
\EndFunction
\end{algorithmic}
\end{algorithm}

\begin{algorithm}
\caption{Computation of determinants using minor recursion}
\begin{algorithmic}[1]
\Function{UpdateMinors}{$\A,\ups,k$}
    \State Assume $\ups_{k+1},\dots,\ups_d$ are known.
    \If{$k=d$}
        \State Set $\det\A_{d,\tau}=(\ups_d)_{j\not\in\tau}$
    \Else
        \State Set $\ups_k=\ups$ and calculate for every suitable $\tau$ the values $\det\A_{k,\tau}$ according to \eqref{eq:def-minors}.
    \EndIf
\EndFunction
\end{algorithmic}
\end{algorithm}

\begin{algorithm}
\caption{IterativeVolume($i,d,D,\tilde\cV,f,\cN_0,\A$)}
\begin{algorithmic}[1]
\Function{IntegrateCell}{$i,\cM,f$}
\State Let $\cN^{(i)}$ be the $K_i$ neighbors of cell $i$.
\State Setup $\A$, initialize it with $\ups_1,\dots,\ups_d=0$.
\State Call \Call{IterativeVolume}{$i,d,D,\tilde\cV,f,\cN^{(i)},\A$}
\EndFunction
\\
\Function{IterativeVolume}{$i,d,D,\tilde\cV,f,\cN_0,\A$}
    \If{$D=1$}
        \State create empty lists $\cV^{(j)}$ for $j\in\cN_0$
        \For{each $(\bsigma,\br)\in\tilde\cV$}
            \State Interpret $\bsigma=(\sigma_k)_k$
            \For{every $\sigma_k$}
                \State push $(\bsigma,\br)$ to $\cV^{(\sigma_k)}$
            \EndFor
        \EndFor
        \State set $\tilde V=0$ and $\tilde F=0$
        \For{each $j\in\cN_0$ with $j>i$}
            \State $V,F=$ \Call{IterativeVolume}{$i,d,D+1,\tilde\cV^{(j)},f,\cN_0\setminus\{j\},\A$}
            \State Set $V=V*\frac{1}{d!}$, $F=F*\frac{1}{d!}$ and update $\tilde V:=\tilde V+V$, $\tilde F:=\tilde F+F$
            \State Let $h$ be the distance of $x_i$ to the plane given by $\tilde \cV^{(j)}$
            \State Define $A^{(i)}(j):=V*\frac dh$, $F^{(i)}(j):=F*\frac dh$
        \EndFor
        \For{each $j\in\cN_0$ with $j<i$}
            \State Let $h$ be the distance of $x_i$ to the mid-plane of $x_i$ and $x_j$
            \State Add: $\tilde V=\tilde V+\frac hdA^{(j)}(i)$, $\tilde F=\tilde F+\frac hdF^{(j)}(i)$
        \EndFor
        \State update: $\tilde F=\frac{d}{d+1}\tilde F+\frac{1}{d+1}f(x_i)\tilde V$
        \State return $\tilde V,\tilde F, A^{(i)},F^{(i)}$
    \EndIf
    \If{$1<D<d$}
        \State create empty lists $\cV^{(j)}$ for $j\in\cN_0$
        \State calculate midpoint $\br_m$ of all $(\bsigma,\br)\in\tilde\cV$ and call \Call{UpdateMinors}{$\A,\br_m,d-D+1$}
        \For{each $(\bsigma,\br)\in\tilde\cV$}
            \State Interpret $\bsigma=(\sigma_k)_k$
            \For{every $\sigma_k$}
                \If{$\sigma_k\in\cN_0$}
                    \State push $(\bsigma,\br)$ to $\cV^{(\sigma_k)}$
                \EndIf
            \EndFor
        \EndFor
        \For{each $j\in \cN_0$}
            \State $V_j,F_j=$\Call{IterativeVolume}{$i,d,D+1,\tilde\cV^{(j)},f,\cN_0,\A$}
        \EndFor
        \State calculate $\tilde V$ and $\tilde F$ as sums of $V_j$ and $F_j$ for each $j\in\cN_0$
        \State return $\tilde V$, $\frac{d-D+1}{d-D+2}\tilde F+\frac1{d-D+2}f(r_m)\tilde V$
    \EndIf
    \If{$D=d$}
        \State take two elements $(\bsigma,\br),(\tilde\bsigma,\tilde\br)\in\tilde\cV$ and call \Call{UpdateMinors}{$\A,\br,2$}
        \State then call $V=$ \Call{UpdateMinors}{$\A,\tilde\br,1$}
        \State return $V,\,\frac12(f(\br)+f(\tilde \br))V$
    \EndIf
\EndFunction
\end{algorithmic}
\end{algorithm}

\end{document}